\documentclass[a4paper, 11pt]{article}  

\usepackage[utf8]{inputenc}
\usepackage[T1]{fontenc}
\usepackage[francais,english]{babel}
\usepackage[top=3.9cm,bottom=4.6cm,left=2.9cm,right=2.9cm]{geometry}
\usepackage{soul}
\usepackage{ulem}
\usepackage{bm}
\usepackage{lmodern}
\usepackage{amsmath}
\usepackage{amssymb}
\usepackage{mathrsfs}
\usepackage{stmaryrd}
\usepackage{graphicx} 
\usepackage{float}
\usepackage{enumitem}
\usepackage{setspace}
\usepackage{multicol}
\usepackage{indentfirst}
\usepackage{xifthen}
\usepackage[hidelinks]{hyperref}
\usepackage{cleveref}
\usepackage{cite}
\usepackage{tikz}
\usepackage{empheq}

\renewcommand{\d}[1][]{\ifthenelse{\isempty{#1}}{\ensuremath{\text{d}}}{\ensuremath{\text{d}^{#1}}}}
\newcommand{\D}{\ensuremath{\mathscr{D}}}

\title{\vspace{10mm} New supersymmetric index of heterotic compactifications with torsion} 

\author{
Dan Isra\"el\thanks{israel@lpthe.jussieu.fr} \ \ and \ Matthieu Sarkis\thanks{msarkis@lpthe.jussieu.fr}\\[2mm]
\small Sorbonne Universit\'es, UPMC Univ Paris 06, UMR 7589, LPTHE, F-75005, Paris, France \\ 
\small CNRS, UMR 7589, LPTHE, F-75005, Paris, France
\vspace{-5mm}
}
\date{}

\begin{document}

\renewcommand{\contentsname}{Table of contents}
\maketitle

\begin{abstract}

We compute the new supersymmetric index of a large class of $\mathcal{N}=2$ heterotic compactifications 
with torsion, corresponding to principal two-torus bundles over warped K3 surfaces with H-flux. 
Starting from a UV description as a (0,2) gauged linear sigma-model with torsion, we use supersymmetric localization 
techniques to provide an explicit expression of the index as a sum over the Jeffrey-Kirwan residues of the one-loop determinant. 
We finally propose a geometrical formula that gives the new
supersymmetric index in terms of bundle data, regardless of any
particular choice of underlying two-dimensional theory.

\end{abstract}

\newpage
\tableofcontents
\newpage

%%%%%%%%%%%%%%%%%%%%%%%%%%%%%%%%%%%%%%%%%%%%%%%%%%%%%%%%%%%%%%%%%%%%%%%%%%%%%%%%%%%%%%%%%%%%%%%%%%%%%%%%%%%%%%%%%%%%%%%%%%%%%%%%%%%%%%%%%%%%%%%%%%%%%%%%%%%%%%%%%%%%%%%%%%%%%%%%%%%%%%%%%%%%%%%%%%%%%%%%%%%%%%%%%%%%%%%%%%%%%%%%%%%%%%%%%%%%%%%%%%%%%%%%%%%%%%%%%%%%%%%%%%%%%%%%%%%%%%%%%%%%%%%%%%%%%%%%%%%%%%%%%%%%%%%%%%%%%%%%%%%%%%%%%%%%%%%%%%%%%%%%%%%%%%%%%%%%%%%%%%%%%%%%%%%%%%%%%%%%%%%%%%%%%%%%%%%%%%%%%%%%%%%%%%%%%%%%%%%%%%%%%%%%%%%%%%%%%%%%%%%%%%%%%%%%%%%%%%%%%%%%%%%%%%%%%%%%%%%%%%%%%%%%%%%%%%%%%%%%%%%%%%%%%%%%%%%%%%%%%%%%%%%%%%%%%%%%%%%%%%%%%%%%%%%%%%%%%%%%%%%%%%%%%%%%%%%%%%%%%%%%%%%%%%%%%%%%%%%%%%%%%%%%%%%%%%%%%%%%%%%%%%%%%%%%%%%%%%%%%%%%%%%%%%%%%%%%%%%%%%%%%%%%%%%%%%%%%%%%%%%%

\section{Introduction}
	
Heterotic compactifications play a fundamental role in building phenomenologically relevant models for particle physics; in this perspective,  
some  realistic Calabi-Yau (CY) models from the point of view of the field content and the interactions were built, see $e.g.$~\cite{Braun:2005ux}. 
However remains the problem of stabilizing the massless moduli characterizing the Calabi--Yau manifold and the stable holomorphic  vector bundle. 
Although a suitable choice of gauge bundle can stabilize a significant fraction of the complex structure moduli~\cite{Anderson:2010mh}, torsional 
compactifications, in which a non-trivial Kalb-Ramond H-flux is turned on, constitute an essential approach towards solving completely this moduli problem. 

The supersymmetry conditions at order $\alpha'$ for $\mathcal{N}=1$ compactifications to four dimensions with H-flux have been known for 
almost thirty years~\cite{Strominger:1986uh}. Nevertheless our knowledge of solutions of these equations, known as Strominger's system, is very 
limited. Indeed the compactification manifold is conformally balanced instead of K\"ahler, and the Bianchi identity, 
see eq.~(\ref{4}), is notoriously  hard to solve as it is non-linear in the flux.\footnote{Only is some specific cases, 
such as the one discussed in this article, does the Bianchi identity boil down to a partial differential equation for a single function, which 
makes the problem simpler to solve.}

There are two pitfalls that await any attempt to construct compactifications with torsion from a low-energy perspective. First, the 
Bianchi identity implies that, if H-flux is present  at leading order in $\alpha'$, there exists no large-volume limit of the 
compactification in general (see~\cite{Melnikov:2014ywa} for a recent discussion). Second, the underlying $(0,2)$ non-linear sigma-model 
is generically destabilized by worldsheet instantons~\cite{Dine:1986zy}. A promising approach that was developed recently is to obtain 
the worldsheet theory as the infrared fixed point of a $(0,2)$ gauge theory, generalizing the well-known Calabi-Yau gauged linear sigma-models 
(GLSMs)~\cite{1993NuPhB.403..159W} to torsion gauged linear sigma-models 
(TGLSMs)~\cite{Adams:2006kb,Adams:2009av,2011JHEP...03..045A,Blaszczyk:2011ib,Quigley:2011pv,Quigley:2012gq,Adams:2012sh}. Worldsheet instanton corrections 
may indeed cancel for theories with a UV (0,2) GLSM description~\cite{Beasley:2003fx}.\footnote{However, some caveats in the arguments of \cite{Beasley:2003fx} concerning 
the absence of worldsheet instantons destabilizing  $(0,2)$ GLSMs have been recently uncovered in~\cite{Bertolini:2014dna}.} 

There exists a single well-known class of compactifications with torsion, given by  principal two-torus bundles over a warped K3 
base together with the pullback of a stable holomorphic vector bundle over K3; following the general usage, they will be named 
Fu-Yau (FY)  compactifications thereafter. These solutions were first obtained by Dasgupta, Rajesh and Sethi 
from type IIB orientifolds by S-duality~\cite{Dasgupta:1999ss}, and subsequently studied geometrically by Goldstein and 
Prokushkin in~\cite{Goldstein:2002pg}, where their $SU(3)$ structure was made explicit. 
Fu and Yau managed to solve the Bianchi identity in~\cite{Fu:2006vj}, using the Chern connection 
(with a sequel \cite{Becker:2006et} discussing more physical aspects), while a different choice of connection 
was put forward in~\cite{Becker:2009df}. These compactifications lead to $\mathcal{N}=2$ or $\mathcal{N}=1$ 
supersymmetry in space-time. The first class of torsion GLSM that was obtained by Adams and 
collaborators~\cite{Adams:2006kb} was especially designed to give a worldsheet theory for the former. 

The microscopic description of Fu-Yau compactifications as torsion GLSMs provides some evidence for their consistency at the quantum level, 
beyond the supergravity regime.\footnote{As for ordinary GLSMs the arguments leading to the absence of destabilization by worldsheet instantons in torsion GLSMs  
should be taken with a grain of salt. In the present case however space-time $\mathcal{N}=2$ supersymmetry presumably prevents such corrections from  
contributing to the effective superpotential.} 
This approach was also used in~\cite{2011JHEP...03..045A} to compute their massless spectra using Landau-Ginzburg 
methods, and in~\cite{2013JHEP...11..093I} to obtain exact statements about their duality symmetries. A very interesting aspect 
of the latter work, which will play an important role in the present paper, was that covariance of the theory under perturbative 
$O(2,2;\mathbb{Z})$ dualities along the two-torus fiber imposes that its  moduli are those of a $c=2$ rational conformal field theory. 

It is natural to ask whether there are other important results regarding heterotic compactifications with torsion that can be obtained in this 
GLSM framework. Typically, the non-conformal gauged linear sigma-models allow one to compute exactly quantities that are invariant under the 
RG flow on the worldsheet. A good example of this is the elliptic genus~\cite{Witten:1986bf} which was  indeed obtained, for CY compactifications, 
using  their formulation as a GLSM~\cite{Gadde:2013dda,2014LMaPh.104..465B,2015CMaPh.333.1241B} and supersymmetric localization~\cite{Pestun:2007rz}. 

In the case of torsion GLSMs for $\mathcal{N}=2$ Fu-Yau compactifications  that interest us in this work, the elliptic genus vanishes since they 
have too many fermionic zero modes.  We will consider instead their {\it new supersymmetric index}~\cite{1992NuPhB.386..405C}, which contains 
important information about four-dimensional physics. It counts the BPS states in space-time~\cite{1996NuPhB.463..315H}, and allows to compute 
the one-loop threshold corrections to the gauge and gravitational couplings of $\mathcal{N}=2$ heterotic compactifications, 
see~\cite{1992NuPhB.383...93A,Antoniadis:1992rq} and $e.g.$~\cite{1996NuPhB.482..187H,Stieberger:1998yi} for subsequent work. While 
these are well-established results for $\text{K3}\times T^2$ compactifications, our main motivation is to extend this analysis to 
the more general case of Fu-Yau geometries.\footnote{This has been done for local models of these compactifications in~\cite{Carlevaro:2012rz} 
by one of the authors, using their solvable conformal field theory description found in~\cite{Carlevaro:2008qf}.}  
			
In this article, we will derive the new supersymmetric index directly from torsion GLSMs corresponding to Fu-Yau compactifications with $\mathcal{N}=2$ 
supersymmetry, using supersymmetric localization. Several steps of the derivation are similar to the computation of the 
elliptic genera for 'ordinary' gauged linear sigma-models~\cite{Gadde:2013dda,2014LMaPh.104..465B,2015CMaPh.333.1241B}. There are however 
important new aspects related to the presence of gauge anomalies canceled against classically non gauge-invariant interactions. 
With the choice of supercharge $\mathcal{Q}$ appropriate to the problem, the action of the torsion multiplet, representing the torus fiber, 
is not $\mathcal{Q}$-exact, and the measure in field space is not $\mathcal{Q}$-invariant; as we will demonstrate, supersymmetric 
localization makes sense nonetheless for the full theory. 

Independently of physics, the elliptic genus of a holomorphic vector bundle over a compact complex manifold is obtained as 
the holomorphic Euler characteristic of a formal power series with vector bundle coefficients; whenever suitable topological 
conditions are met (essentially the tadpole conditions), it gives a weak Jacobi form. The twining partition function 
that we define as an intermediate step in the computation of  the new supersymmetric index, see eqns~\eqref{eq:index} and~\eqref{eq:FYpart}, 
is the natural (non-holomorphic) generalization of the Calabi-Yau elliptic genus to Fu-Yau geometries. 
More generally, it provides a non-holomorphic genus for  principal two-torus bundles over CY $d$-folds, which transforms as a 
weak Jacobi form,  and whose topological nature follows from quantization of the torus moduli. We will present a definition of this 
quantity in geometrical terms, independently of any GLSM or other worldsheet formulation, in eq.~(\ref{eq:nonholgen}); a proof of this 
statement will be provided for an example based on the quartic. 
			
The plan of this article is as follows. 
In section~\ref{sec:GLSM} we review the construction of torsion GLSMs. Then in section~\ref{sec:indexdef} we present the 
new supersymmetric index, and proceed in section~\ref{sec:local} to its path integral computation using localization. In 
section~\ref{sec:examples} we generalize the results to higher rank gauge groups, provide an anomaly-free charge assignment for 
a large class of models and illustrate our results with an explicit example. 
In section~\ref{sec:geom} we provide the geometrical formula for the new supersymmetric index, and finally  we summarize 
the work exposed in this article in section~\ref{sec:conc} and give directions for future work. Conventions for $(0,2)$ 
superspace are gathered in \cref{appendixConventions}, some results about theta functions and modular forms can be found in~\cref{appendixTheta}, 
and a summary of Fu-Yau geometry is given in~\cref{appendixFuYau}.
		
	\bigskip
	\noindent \textit{\underline{Conventions}:}
	\begin{itemize}
		\item In the following, we set $\alpha'=1$, which means that the self-dual radius is one.
		\item The real part (resp. imaginary part) of any complex quantity is denoted by an index 1 (resp. 2).
		\item $4\pi S=\int\d[2]z\ \mathcal{L}$.
		\item left-moving $\leftrightarrow$ holomorphic.
		\item One defines $q=\exp (2i\pi\tau)$ and $w=\exp (2i\pi y)$.
		\item The volume of the worldsheet torus is $\int\d[2]z=2\tau_{2}$.
	\end{itemize}

%%%%%%%%%%%%%%%%%%%%%%%%%%%%%%%%%%%%%%%%%%%%%%%%%%%%%%%%%%%%%%%%%%%%%%%%%%%%%%%%%%%%%%%%%%%%%%%%%%%%%%%%%%%%%%%%%%%%%%%%%%%%%%%%%%%%%%%%%%%%%%%%%%%%%%%%%%%%%%%%%%%%%%%%%%%%%%%%%%%%%%%%%%%%%%%%%%%%%%%%%%%%%%%%%%%%%%%%%%%%%%%%%%%%%%%%%%%%%%%%%%%%%%%%%%%%%%%%%%%%%%%%%%%%%%%%%%%%%%%%%%%%%%%%%%%%%%%%%%%%%%%%%%%%%%%%%%%%%%%%%%%%%%%%%%%%%%%%%%%%%%%%%%%%%%%%%%%%%%%%%%%%%%%%%%%%%%%%%%%%%%%%%%%%%%%%%%%%%%%%%%%%%%%%%%%%%%%%%%%%%%%%%%%%%%%%%%%%%%%%%%%%%%%%%%%%%%%%%%%%%%%%%%%%%%%%%%%%%%%%%%%%%%%%%%%%%%%%%%%%%%%%%%%%%%%%%%%%%%%%%%%%%%%%%%%%%%%%%%%%%%%%%%%%%%%%%%%%%%%%%%%%%%%%%%%%%%%%%%%%%%%%%%%%%%%%%%%%%%%%%%%%%%%%%%%%%%%%%%%%%%%%%%%%%%%%%%%%%%%%%%%%%%%%%%%%%%%%%%%%%%%%%%%%%%%%%%%%%%%%%%%%

\section{Gauged linear sigma-models with torsion}
\label{sec:GLSM}

We review in this section the construction of torsion gauged linear sigma-models proposed in~\cite{Adams:2006kb}. 
These are gauge theories in two-dimensions with $(0,2)$ supersymmetry which are expected to flow in the infrared to $(0,2)$ 
non-linear sigma-models whose target space corresponds to Fu-Yau compactifications; a brief presentation of these 
non-K\"ahler heterotic solutions is given in appendix~\ref{appendixFuYau}.

As the first step of this construction, one considers a standard (0,2) gauged linear sigma-model for the K3 base; 
generically such model suffers from gauge anomalies, that, in the usual case of Calabi-Yau GLSMs, should be made 
to vanish by a suitable choice of field content hence of gauge bundle in space-time. In the present case, one cancels instead the 
anomalous variation of the functional measure against a classically non-gauge-invariant Lagrangian for a {\it torsion multiplet} 
modeling the $T^{2}$ principal bundle, thereby realizing the Green-Schwarz mechanism on the worldsheet.  

\subsection{Anomalous gauged linear sigma-model for the base}

For simplicity of the discussion, we restrict ourselves in the following discussion to the case of a $U(1)$  gauge group on the worldsheet;  
the generalization to higher rank gauge groups is rather straightforward and will be 
briefly mentioned in section~\ref{sec:examples}. The conventions we use for $(0,2)$ superfields, as well as the 
components Lagrangian, are given in appendix~\ref{appendixConventions}. 
			
A $(0,2)$ gauged linear sigma-model for a complete intersection Calabi-Yau manifold in a weighted projective 
space~\cite{1993NuPhB.403..159W} contains first a set of $n$ $(0,2)$ chiral multiplets $\Phi_i$, as well as a set 
of $p$ Fermi multiplets $\tilde{\Gamma}_\alpha$, interacting through the superpotential 
\begin{equation}
\mathcal{L}_\textsc{t} =\int\d\theta^{+}\, \tilde{\Gamma}_{\alpha}G^{\alpha}(\Phi_i) + h.c. \, ,
\end{equation}
where the $G^\alpha (\phi_i)$ are quasi-homogeneous polynomials of the appropriate degree to preserve gauge 
invariance at the classical level and, geometrically, to obtain a hypersurface of vanishing first Chern class. This Calabi-Yau 
hypersurface corresponds then to the complete intersection $\bigcap_{\alpha=1}^p \left\{\phi_i\ |\ G^{\alpha}(\phi_i)=0\right\}$.

Second, the holomorphic vector bundle is described, in the 
simplest examples, by adding a set of $s+1$ Fermi multiplets $\Gamma_a$, a single chiral multiplet $P$ and the superpotential 
\begin{equation}
\mathcal{L}_\textsc{v} =\int\d\theta^{+}\, P \, \Gamma_{a} \, J^{a}(\Phi_i)  + h.c.\, ,
\end{equation}
where the $J^a$ are again quasi-homogeneous polynomials.  Let us denote the gauge charges of the different superfields as 
(with $Q_P$ and $Q_\alpha$ negative, the other ones positive):
\begin{equation}
{\setlength\arraycolsep{8pt}
\begin{array}{|c||c|c|c|c|}
\hline
& \Phi_{i} & P & \tilde{\Gamma}_{\alpha} & \Gamma_{a} \\
\hline\hline U(1)_{\text{gauge}} & Q_{i} & Q_{P} & Q_{\alpha} & Q_{a} \\
\hline
\end{array}
}
\end{equation}
In the geometrical "phase", where the real part of the Fayet-Iliopoulos coupling is taken large and positive, 
one expects that the model flows to a non-linear sigma-model on 
the CY hypersurface with left-handed fermionic degrees of freedom transforming as sections of a rank $s$ holomorphic vector 
bundle $\mathcal{V}$, determined by the short exact sequence\footnote{For simplicity of the presentation we do not consider adding 
fermionic gaugings. For more details, see for instance the review~\cite{1995hepc.conf..322D}.}
\begin{equation}
0 \longrightarrow \mathcal{V} \stackrel{\iota}{\longrightarrow} \bigoplus_{a=1}^{s+1} \mathcal{O} (Q_a) \stackrel{\otimes J^a}{\longrightarrow} 
\mathcal{O} (-Q_P) \longrightarrow 0\, .
\end{equation}

As the $(0,2)$ multiplets contain chiral fermions there are potentially gauge anomalies on the worldsheet 
that should be canceled. The model should also contain a non-anomalous global right-moving $U(1)$ symmetry which 
corresponds in the infrared to the $U(1)_{\textsc{r}}$ symmetry of the  $N=2$ superconformal algebra, and a global 
left-moving $U(1)_{\textsc{l}}$ symmetry, used to implement the left-moving GSO projection.

The variation of the effective Lagrangian under a super-gauge transformation of chiral parameter $\Xi$ writes
\begin{equation}
\delta_{\Xi}\mathcal{L}_{\text{eff}}=-\frac{\mathcal{A}}{4}\int\d\theta^{+}\ \Xi  \Upsilon\ +h.c.,
\end{equation}
with $\Upsilon$ the field strength superfield, and the anomaly coefficient 
\begin{equation}
\mathcal{A}=\sum_{i}Q_{i}^{\ 2}+Q_{P}^{\ 2}-\sum_{\alpha}Q_{\alpha}^{\ 2}-\sum_{a}Q_{a}^{\ 2},
\label{Anomaly}
\end{equation}
which measures the difference between the second Chern character of the 
tangent bundle of the base manifold and the second Chern character of the vector bundle over the latter. If one considers a model 
with $\mathcal{A}\neq 0$, then the theory is at this point ill-defined quantum mechanically. 

\subsection{Two-torus principal bundle and anomaly cancellation}
	
In the original work of Adams and collaborators~\cite{Adams:2006kb}, the two-torus bundle over the K3 base is built up 
by first constructing a $\mathbb{C}^{*}\times\mathbb{C}^{*}$ non-compact bundle, and then changing complex structure in field 
space, allowing to discard the decoupled non-compact part from the $\mathbb{C}^{*}\times\mathbb{C}^{*}=\mathbb{C}\times T^{2}$ bundle, 
while preserving $(0,2)$ supersymmetry.

To start, one introduces two extra chiral multiplets $\Omega_{1}=(\omega_{1},\chi_{1})$ and 
$\Omega_{2}=(\omega_{2},\chi_{2})$,  whose (imaginary) shift symmetry is gauged as 
\begin{equation}
\delta_{\Xi} \, \Omega_{\ell} = -i\, M_\ell\, \Xi \ , \quad M_\ell \in \mathbb{Z}\ , \quad \ell=1,2\, .
\end{equation}
The compact bosonic fields $\text{Im} (\omega_\ell)$ will ultimately parametrize the torus fiber. 

A generic two-torus is characterized by a complex structure $T=T_1+iT_2$ and a complexified K\"ahler modulus 
$U=U_1+iU_2$, such that the metric and Kalb-Ramond field are given by
\begin{equation}
\label{eq:torusmod}
G=\frac{U_{2}}{T_{2}}\begin{pmatrix}1&T_{1}\\T_{1}&|T|^{2}\end{pmatrix}\ , \quad
B=\begin{pmatrix}0&U_{1}\\-U_{1}&0\end{pmatrix}.
\end{equation}

The Lagrangian for $\Omega_1$ and $\Omega_2$, corresponding 
to a complexification of this two-torus, reads~\cite{2013JHEP...11..093I}:
{\allowdisplaybreaks
\begin{align}
\mathcal{L}_{s}=&-\frac{iU_{2}}{4T_{2}}\int\d[2]\theta\ \left(\Omega_{1}+\bar{\Omega}_{1}+T_{1}\left(\Omega_{2}+\bar{\Omega}_{2}\right)
+2(M_{1}+T_{1}M_{2})A_{+}\right)\times\nonumber\\
&\ \ \ \ \ \ \ \ \ \ \ \ \ \ \times\left(\partial_{-}\left(\Omega_{1}-\bar{\Omega}_{1}+T_{1}\left(\Omega_{2}-\bar{\Omega}_{2}\right)\right)
+2i(M_{1}+T_{1}M_{2})A_{-}\right)\nonumber\\
&-\frac{iU_{2}T_{2}}{4}\int\d[2]\theta\ \left(\Omega_{2}+\bar{\Omega}_{2}+2M_{2}A_{+}\right)
\left(\partial_{-}\left(\Omega_{2}-\bar{\Omega}_{2}\right)+2iM_{2}A_{-}\right)\nonumber\\
&+\frac{iU_{1}}{4}\int\d[2]\theta\ \left\{\left(\Omega_{1}+\bar{\Omega}_{1}+2M_{1}A_{+}\right)
\left(\partial_{-}\left(\Omega_{2}-\bar{\Omega}_{2}\right)+2iM_{2}A_{-}\right)-\right.\nonumber\\
&\ \ \ \ \ \ \ \ \ \ \ \ \ \ \left.-\left(\Omega_{2}+\bar{\Omega}_{2}+2M_{2}A_{+}\right)\left(\partial_{-}\left(\Omega_{1}-
\bar{\Omega}_{1}\right)+2iM_{1}A_{-}\right)\right\}\nonumber\\
&-\frac{iN^{i}}{2}\int\d\theta^{+}\, \Upsilon\, \Omega_{i}+h.c. 
\end{align}
}
The couplings between the chiral superfields $\Omega_{\ell}$ and the field strength superfield $\Upsilon$ contain field-dependent 
Fayet-Iliopoulos (FI) terms (last line) that are classically non-invariant under (super)gauge transformations:
\begin{equation}
\delta_\Xi \mathcal{L}_{s} = -\frac{N^i M_i}{2} \int \d\theta^{+}\, \Upsilon\,  \Xi + h.c.\, .
\end{equation}
This gauge  variation should be such that it compensates the one-loop anomaly~\eqref{Anomaly} 
of the base GLSM; this can be viewed as a worldsheet incarnation of the Green-Schwarz mechanism. 
Finally, in order for the action to be single-valued under $\omega_{i}\sim\omega_{i}+2i \pi$ 
in any instanton sector, the couplings 
$N^{i}$ should be integer-valued. 

\subsubsection*{Moduli quantization}	
	
In order to restrict the non-compact $\mathbb{C}^{*}\times\mathbb{C}^{*}$ fibration described above to a $T^{2}$ fibration 
while maintaining  $(0,2)$ worldsheet supersymmetry, one has to define a complex structure in field space that allows for 
a decoupling of the real part of these multiplets. This is compatible with supersymmetry provided that the couplings between the gaugini and 
the fermionic components of the superfields $\Omega_i$ vanish~\cite{Adams:2006kb}. It amounts to the following 
relations between the Fayet-Iliopoulos parameters and the charges~\cite{2013JHEP...11..093I}
\begin{subequations}
\label{eq:condcharges}
\begin{align}
N^{1}&=-\frac{U_{2}}{T_{2}}\text{Re}(M)-U_{1}M_{2}\ \in\mathbb{Z}\, ,\\
N^{2}&=-\frac{U_{2}}{T_{2}}\text{Re}(\bar{T}M)+U_{1}M_{1}\ \in\mathbb{Z}\, ,
\end{align}
\end{subequations}
with the complex charge $M$ defined as 
\begin{equation}
\label{eq:Mdef}
M=M_{1}+TM_{2}\, .
\end{equation} 
Using these relations the gauge-variation of the field-dependent Fayet-Iliopoulos term reads:
\begin{equation}
\label{eq:deltator}
\delta_\Xi \mathcal{L}_{s} = \frac{U_2}{2T_2} |M|^2 \int \d\theta^{+}\, \Upsilon  \Xi + h.c. \, ,
\end{equation}
which should be cancelled against the gauge anomaly from the chiral fermions in 
order to get a consistent quantum theory. One obtains the condition  
\begin{equation}
\label{eq:tadpole}
\sum_{i}Q_{i}^{\ 2}+Q_{P}^{\ 2}-\sum_{\alpha}Q_{\alpha}^{\ 2}-\sum_{a}Q_{a}^{\ 2}-\frac{2U_{2}}{T_{2}}|M|^{2}=0\,  ,
\end{equation}
reproducing the tadpole condition from the integrated Bianchi identity in Fu-Yau compactifications~\cite{Becker:2006et}, see 
app.~\ref{appendixFuYau}.

The torus moduli $T$ and $U$ are partially quantized by the pair of supersymmetry 
conditions~(\ref{eq:condcharges}); in a  model with worldsheet gauge 
group $U(1)^k$ one obtains one such condition for each complex charge $M_\kappa$, 
hence the moduli are generically fully quantized. As was shown in~\cite{2013JHEP...11..093I}, 
covariance of the theory under T-duality symmetries along the fiber provides another way of understanding 
quantization of the torus moduli. Under the transformation $U\mapsto -1/U$ in $PSL(2;\mathbb{Z})_U$, each 
complex charge $M$ is mapped to $-\bar{U} M$. For consistency this charge should 
belong to the same lattice as the original one, namely 
$\bar{U} M \in\mathbb{Z}+T\mathbb{Z}$.

Demanding that this property holds for every topological charge in the model is actually a non-trivial 
statement. Generically, this is true if and only if the elliptic curve $E_T=\mathbb{C}/(\mathbb{Z}+T\mathbb{Z})$ admits a non-trivial endomorphism 
\begin{align}
E_T&\rightarrow E_T\nonumber\\
z&\mapsto \bar{U}z \, ,
\end{align}
which is known as {\it complex multiplication}. This property holds if and only if 
both $U$ and $T$ are valued in the same imaginary quadratic number field 
$\mathbb{Q}(\sqrt{D})$ with $D$ the discriminant of a positive definite integral quadratic form:
\begin{equation}
D = b^2 -4ac  <0 \ , \quad a,b,c \in \mathbb{Z} \ , \quad a>0\, .
\end{equation} 
Crucially, $c=2$ conformal field theories with a two-torus target space 
are rational iff their $T$ and $U$ moduli satisfy these conditions~\cite{2004CMaPh.246..181G,Hosono:2002yb}; this 
property will play an important role in section~\ref{sec:local}.

One could also consider incorporating in the torsion GLSM terms corresponding to extra Abelian gauge 
bundles over the total space $\mathcal{M}$  
(that would be Wilson lines along the torus in the $K3\times T^2$ case), which are indeed allowed by the 
space-time supersymmetry constraints~\cite{Becker:2006et}. We leave this generalization of the models for future 
work.\footnote{It may be difficult to incorporate such bundles in an explicit $(0,2)$ GLSM framework; see~\cite{2013JHEP...11..093I} for a 
preliminary discussion about this aspect.}

\subsubsection*{Torsion multiplet}

Whenever the supersymmetry conditions~(\ref{eq:condcharges}) are met, the non-compact real parts of $\Omega_{1,2}$ 
decouple and one can reorganize their imaginary parts into a {\it torsion multiplet} $\Theta=\left(\alpha,\chi\right)$, with 
\begin{align}
\alpha&=\text{Im}\, (\omega_{1})+T\, \text{Im}\, (\omega_{2}),\\
\chi&=\text{Re}\, (\chi_{1})+\bar{T}\, \text{Re}(\chi_{2}),
\end{align}
shifted as $\delta_\Xi\Theta = -M\, \Xi$ under supergauge transformations, with the complex charge $M$ defined in eq.~(\ref{eq:Mdef}). 
The Lagrangian is given (omitting temporarily the topological B-field term for simplicity) by 
\begin{equation}
\mathcal{L}_{\text{t.m.}}=-i\frac{U_{2}}{T_{2}}\int\d[2]\theta\left(\bar{\Theta}+2i\bar{M}A_{+}\right)\mathcal{D}_{-}
\Theta-\frac{\bar{M}}{2}\frac{U_{2}}{T_{2}}\int\d\theta^{+}\Theta \Upsilon \ +h.c.,
\end{equation}
where the superspace covariant derivative reads $\mathcal{D}_{-}\Theta=\partial_{-}\Theta-\frac{iM}{2}(2\partial_{-}A_{+}+iA_{-})$. 
						
As usual going to Wess-Zumino gauge is convenient in order to exhibit the physical degrees of 
freedom; in the present situation one should not forget nevertheless that the theory is not classically 
gauge invariant, hence such gauge choice only makes sense in the path integral of the full quantum theory, 
including the base GLSM, as will be clear below when supersymmetric localization will be put into action. 

In this gauge the torsion multiplet contains a compact complex boson coupled chirally to a gauge 
field and a free right-moving Weyl fermion. After going to Euclidean signature\footnote{One first Wick rotates 
to Euclidean time $\sigma^{2}=-i\sigma^{0}$. Complex coordinates are then 
defined by $z=\sigma^{1}+i\sigma^{2}$ and $\bar{z}=\sigma^{1}-i\sigma^{2}$.} and some rescaling of the fields, one has
\begin{equation}
\frac{T_{2}}{U_{2}}\, \mathcal{L}_{\text{t.m.}}=\nabla_{z}\bar{\alpha}\nabla_{\bar{z}}
\alpha+\nabla_{z}\alpha\nabla_{\bar{z}}\bar{\alpha}-\frac{1}{2}\left(M\bar{\alpha}+\bar{M}\alpha\right)
a_{z\bar{z}}+2\bar{\chi}\partial\chi,
\end{equation}
with $\nabla_{z}\alpha=\partial\alpha+Ma_{z}$ and $\nabla_{\bar{z}}\alpha=\bar{\partial}\alpha+Ma_{\bar{z}}$, 
and where $a_{z\bar{z}}=2\left(\partial a_{\bar{z}}-\bar{\partial}a_{z}\right)$ denotes the field strength of the 
gauge field. After integrating by parts, one gets the following Lagrangian
\begin{equation}
\label{eq:chiralcoupling}
\frac{T_{2}}{U_{2}}\, \mathcal{L}_{\text{t.m.}}=2\partial\bar{\alpha}\bar{\partial}\alpha
+2Ma_{\bar{z}}\partial\bar{\alpha}+2\bar{M}a_{\bar{z}}
\partial\alpha+2|M|^{2}a_{z}a_{\bar{z}}+2\bar{\chi}\partial\chi  + t.d. \, , 
\end{equation}
where the left-moving $U(1)$ currents $\partial \alpha$  and $\partial \bar \alpha$  are coupled to 
the gauge fields, but not the right-moving ones. 

Since we are working in Wess-Zumino gauge, the appropriate supersymmetry transformations are given by  
\begin{equation}
\label{eq:susytransfo}
\delta_{\epsilon}=\left(\epsilon Q_{+}-\bar{\epsilon}\bar{Q}_{+}+\delta_{\text{gauge}}\right),
\end{equation}
where $\delta_{\text{gauge}}$ refers to the supergauge transformation which is needed to restore Wess-Zumino gauge after 
the supersymmetry transformation, corresponding to the chiral parameter $\Xi_{wz}=i\bar{\epsilon}\theta^{+}a_{\bar{z}}$. 
The transformation properties of the different component fields are listed in \cref{appendixConventions}, 
eq.~(\ref{eq:comptrans}); the Lagrangian for the torsion multiplet is not invariant under this transformation, 
but its variation is precisely such that it compensates the variation of the effective action of the base GLSM under the 
gauge transformation back to WZ gauge. 
							
To summarize, a consistent torsion gauged linear sigma-model is given by a base K3 GLSM whose gauge 
anomaly is canceled by a torsion multiplet, provided that the tadpole condition~(\ref{eq:tadpole}) 
holds. Finally one has to choose the $U(1)_{\textsc{l}}$ and $U(1)_{\textsc{r}}$ charges of all the multiplets 
in order to cancel the global anomalies, and to obtain the correct central charges of the IR superconformal algebra and 
rank of the vector bundle. The global charges of the torsion multiplet, proportional to their gauge charge $M$, correspond 
naturally to charges under a shift symmetry.  Consistent choices of global charges will be given in section~\ref{sec:examples}.

%%%%%%%%%%%%%%%%%%%%%%%%%%%%%%%%%%%%%%%%%%%%%%%%%%%%%%%%%%%%%%%%%%%%%%%%%%%%%%%%%%%%%%%%%%%%%%%%%%%%%%%%%%%%%%%%%%%%%%%%%%%%%%%%%%%%%%%%%%%%%%%%%%%%%%%%%%%%%%%%%%%%%%%%%%%%%%%%%%%%%%%%%%%%%%%%%%%%%%%%%%%%%%%%%%%%%%%%%%%%%%%%%%%%%%%%%%%%%%%%%%%%%%%%%%%%%%%%%%%%%%%%%%%%%%%%%%%%%%%%%%%%%%%%%%%%%%%%%%%%%%%%%%%%%%%%%%%%%%%%%%%%%%%%%%%%%%%%%%%%%%%%%%%%%%%%%%%%%%%%%%%%%%%%%%%%%%%%%%%%%%%%%%%%%%%%%%%%%%%%%%%%%%%%%%%%%%%%%%%%%%%%%%%%%%%%%%%%%%%%%%%%%%%%%%%%%%%%%%%%%%%%%%%%%%%%%%%%%%%%%%%%%%%%%%%%%%%%%%%%%%%%%%%%%%%%%%%%%%%%%%%%%%%%%%%%%%%%%%%%%%%%%%%%%%%%%%%%%%%%%%%%%%%%%%%%%%%%%%%%%%%%%%%%%%%%%%%%%%%%%%%%%%%%%%%%%%%%%%%%%%%%%%%%%%%%%%%%%%%%%%%%%%%%%%%%%%%%%%%%%%%%%%%%%%%%%%%%%%%%%%%%
		
\section{New supersymmetric index of \texorpdfstring{$\mathbf{\mathcal{N}=2}$}{N=2}  compactifications}
\label{sec:indexdef}

We consider $\mathcal{N}=2$ compactifications to four dimensions of the $E_8 \times E_8$ heterotic string 
theory. For any $(0,2)$ superconformal field theory with $(c,\bar{c})=(22,9)$ corresponding to the 'internal' 
degrees of freedom of such compactification, the new supersymmetric index is defined as the following 
trace over the Hilbert space in the right Ramond sector:
\begin{equation}
\label{eq:newdef}
Z_\textsc{new} (\tau,\bar \tau) = 
\frac{1}{\eta(\tau)^2}\text{Tr}_\textsc{r}\left[  \bar{J}_0 (-1)^{F_R} q^{L_0-c/24}\bar{q}^{\bar{L}_0-\bar{c}/24} \right]\, ,
\end{equation}
where $F_R$ is the right-moving fermion number and $\bar{J}_0$ the zero-mode of the right-moving R-current, 
which is part of the (right-moving) superconformal algebra. In general, this index is independent of D-term deformations, 
while it is sensitive to F-term deformations~\cite{1992NuPhB.386..405C}.

It was observed in~\cite{1992NuPhB.383...93A,Antoniadis:1992rq} that the threshold corrections to the 
gauge and gravitational couplings of $\mathcal{N}=2$ heterotic string compactifications on $\text{K3}\times T^2$  are 
easily obtained in terms of the new supersymmetric index~(\ref{eq:newdef}).  
Furthermore Harvey and Moore showed in~\cite{1996NuPhB.463..315H} that  it counts the four-dimensional  BPS states as 
\begin{equation}
\label{eq:BPSsum}
-\frac{1}{2i \eta^2}Z_\textsc{new} (q,\bar q) = \sum_{\text{BPS vectors}}q^\Delta \bar{q}^{\bar \Delta}-\sum_{\text{BPS hypers}} 
q^\Delta \bar{q}^{\bar \Delta}\, .
\end{equation}
One of the goals of this paper is to extend this analysis to Fu-Yau geometries. Formula~(\ref{eq:BPSsum}) was proven using representation 
theory of the $\mathcal{N}=4$ superconformal algebra underlying the $K3$ CFT. As was explained in~\cite{Melnikov:2010pq}, 
non-linear sigma-models with a Fu-Yau target space are invariant under the action of the generators of a 
$(0,2) \oplus (0,4)$ superconformal algebra, at the classical level,  hence we expect that a similar reasoning holds 
in the present case.

\subsection{New supersymmetric index of \texorpdfstring{K3 $\times$ T$^\mathbf{2}$ }{K3*T2} compactifications}

We first review the computation of the new supersymmetric index in the familiar case of $\text{K3}\times T^2$ compactifications of the 
$E_8\times E_8$ heterotic string, without Wilson lines for simplicity. We emphasize the role of the left-moving GSO projection 
and the formulation of the index as a chiral orbifold in order to facilitate the generalization to Fu-Yau compactifications in the 
next subsection. 

We assume that the gauge bundle lies in the first $E_8$ only. More specially, we consider a gauge bundle $V$ with the embedding 
$V\subset SO(2n) \subset E_8$. The internal CFT is then the tensor product of a $(0,2)$ theory with $(c,\bar c)=(14,9)$ and 
a  $(c,\bar c)=(8,0)$ theory corresponding to the second $E_8$ factor. 

Using the factorization of the $(c,\bar c)=(14,9)$ CFT in the two-torus and  K3 factors, hence the decomposition of the corresponding 
$(0,2)$ superconformal algebra into the direct sum $(0,2)\oplus (0,4)$, we split the right-moving R-current as follows:
\begin{equation}
\bar{J} = \bar{J}^{\, \textsc{T}^2} + \bar{J}^{\, \text{K3}}\, .
\end{equation}
It allows to expand the superconformal index into the sum of two terms. For the second one, we get 
\begin{equation}
\text{Tr}_\textsc{r}\left[ \bar{J}_{0}^{\, \text{K3}} (-1)^{F_R} q^{L_0-c/24}\bar{q}^{\bar{L}_0-\bar{c}/24} \right]= 0\, ,
\end{equation}
for two different reasons. First, the fermionic partners of the $T^2$ have a pair of fermionic zero modes of opposite fermion numbers, 
hence the trace over the two-torus Hilbert space vanishes. Second the $\text{K3}$ SCFT has $\mathcal{N}=(0,4)$ superconformal symmetry, 
hence the eigenvalues of $\bar{J}_{0}^{\text{K3}}$, which are twice the eigenvalues of the Cartan current of the $SU(2)_1$ 
R-symmetry, come in pairs of opposite sign~\cite{1996NuPhB.463..315H}.

In order to trace over the internal Hilbert space of the theory we have to define a left-moving GSO projection 
corresponding to the first $E_8$ factor. We assume the existence of a  $U(1)_{\textsc{l}}$ left-moving symmetry, 
acting on the $(0,4)$ SCFT describing the K3 surface as on the remaining $(8-n)$ free left-moving Weyl fermions of the first $E_8$.
		
We consider the following twining partition function in the RR sector, with a chemical potential $y$ for this $U(1)_{\textsc{l}}$ symmetry: 
\begin{equation}
Z_{\text{K3}\times T^{2}} (\tau,\bar \tau,y)=\frac{1}{\bar{\eta}(\bar{\tau})^{2}}\text{Tr}_{\textsc{rr},\mathcal{H}_{\text{K3}\times T^{2}}}\left[  e^{2i\pi y J_{0}}\bar{J}_{\, 0}
(-1)^{F} q^{L_0-c/24}\bar{q}^{\bar{L}_0-\bar{c}/24} \right]\, ,\label{eq:FYtrace}
\end{equation}
with $J_{0}$ the left-moving $U(1)$ current, $(-1)^F=\exp i\pi(J_{0}-\bar{J}_{0})$ 
and where the trace is over the Hilbert space of the $\text{K3}\times T^2$ $(0,2)$ superconformal field theory with $(c,\bar c)=(6+n,9)$.  
Then, using standard orbifold formul\ae, see $e.g.$~\cite{Kawai:1994np}, the new 
supersymmetric index is obtained as a sum over the sectors of the chiral $\mathbb{Z}_2$ quotient corresponding 
to the left-moving GSO projection:
\begin{equation}
\label{eq:newGSO}
Z_\textsc{new} (\tau,\bar \tau) = \frac{\bar{\eta}^{2}E_4 (q,0)}{2\eta^{10}} \sum_{\gamma,\delta=0}^1 
q^{\gamma^2} \left.\left\{\left(\frac{\vartheta\left(\tau|y\right)}{\eta(\tau)}\right)^{8-n}Z_{\text{K3}\times T^{2}} (\tau,\bar \tau,y)\right\}\right|_{y=\frac{\gamma\tau+\delta}{2}}\, ,
\end{equation}
where the modular form $E_4(q,0)$ comes from the contribution of the second $E_8$ factor, see \cref{appendixTheta}.  

The partition function over the two-torus degrees of freedom is straightforward. For a torus with complex and K\"ahler moduli 
$T$ and $U$ the soliton sum $\Xi_{2,2} (U,T|\tau,\bar \tau)$ is given by
\begin{equation}
\Xi_{2,2}(\tau,\bar{\tau}|T,U)=\sum\limits_{m_{i},n_{i}\in\mathbb{Z}}
\exp\left(-\frac{\pi}{\tau_2} \frac{U_2}{T_2}\big|m_1 + n_1 \tau + T (m_2 + n_2 \tau)\big|^{2}+2i\pi  U (m_{1}n_{2}-n_{2}m_{1})\right).
\end{equation}
Then the contribution to the partition function~({\ref{eq:FYtrace}) reads		
\begin{multline}
\label{eq:torcont}
\text{Tr}_{\textsc{rr},\mathcal{H}_{T^2}} \left[\bar{J}_0 (-1)^{\bar{J}_0} q^{L_0-c/24}\bar{q}^{\bar{L}_0-\bar{c}/24} \right] 
=\left. \frac{1}{2i\pi} \frac{\partial}{\partial\theta}\right|_{\theta=1/2} 
\text{Tr}_{\textsc{rr},\mathcal{H}_{T^2}} \left[e^{2i\pi \theta \bar{J}_0} q^{L_0-c/24}\bar{q}^{\bar{L}_0-\bar{c}/24} \right] \\
= \frac{\Xi_{2,2} (T,U)}{\eta^2 \bar{\eta}^2} \left.\frac{\partial}{\partial\epsilon}\right|_{\epsilon=0} 
\frac{\vartheta_1(\bar q,e^{2i\pi\epsilon})}{\bar \eta} = \frac{\Xi_{2,2} (T,U)}{i\eta^2}\, .
\end{multline}

Finally one needs to compute the trace over the Hilbert space of the $(0,4)$ theory with a K3 target space. 
Let us consider the case of the standard embedding of the spin connection in the gauge connection, enhancing the supersymmetry 
of the K3 SCFT to $\mathcal{N}=(4,4)$. Then plugging back the expression~(\ref{eq:torcont}) 
into equation~(\ref{eq:newGSO}), and tracing over the Hilbert space of the 6 free Weyl 
fermions with twisted boundary conditions, one gets finally the index in terms of the K3 elliptic genus~\cite{1996NuPhB.463..315H}:
\begin{equation}
\label{eq:newk3t2}
Z_\textsc{new}  = \frac{E_4 (\tau,0)\Xi_{2,2}}{2i\eta^{12}\bar{\eta}^{2}} 
\sum_{\gamma,\delta=0}^1  q^{\gamma^2}
\left(\frac{\vartheta_1 (\tau,\frac{\gamma \tau + \delta}{2})}{\eta(\tau)}\right)^{6} 
Z_{\text{K3}}^{\textsc{ell}} \left(\tau,\frac{\gamma \tau + \delta}{2}\right)\, ,
\end{equation}
where the $(2,2)$ elliptic genus of K3 is defined by 
\begin{equation}
Z_{\text{K3}}^{\textsc{ell}} (\tau,y) = \text{Tr}_{\textsc{rr},\mathcal{H}_{\text{K3}}} \left [ 
e^{2i\pi y J_0} (-1)^F q^{L_0-c/24}\bar{q}^{\bar{L}_0-\bar{c}/24} \right]\,. 
\end{equation} 
As the elliptic genus is a topological invariant~\cite{Witten:1986bf}, it can be computed anywhere in the moduli space of 
K3 compactifications, for instance using Landau--Ginzburg 
orbifolds~\cite{Intriligator:1990ua,Kawai:1993jk,Kawai:1994np}, or toroidal orbifolds~\cite{Schellekens:1986yj}.
It was shown recently~\cite{Gadde:2013dda,2014LMaPh.104..465B,2015CMaPh.333.1241B} how to compute the elliptic genus 
directly at the level of the gauged linear sigma-model, using supersymmetric localization~\cite{Pestun:2007rz}.  
As we shall see in the next section, this localization method can be generalized to compute the new 
supersymmetric index of Fu-Yau compactifications.

The new supersymmetric index of $\text{K3} \times T^2$ compactifications is actually universal, $i.e.$ independent of the choice 
of gauge bundle, as was shown in~\cite{Kiritsis:1996dn}, and reviewed recently in~\cite{2013JHEP...09..030C}. 
The quantity $\tau_2 Z_\textsc{new}$ should be a non-holomorphic modular form of weight -2, with a pole at the infinite cusp 
(this will remain valid in the case of Fu-Yau compactifications). Factorizing the index as 
\begin{equation}
Z_\textsc{new} (\tau, \bar \tau) = -2 i \frac{\Xi_{2,2} (\tau, \bar \tau)}{\eta(\tau)^{4}} \mathcal{G}_{\text{K3}} (\tau)\, ,
\end{equation}
one can show that $\eta^{20} \mathcal{G}_{\text{K3}} (\tau)$ should be a holomorphic modular form of weight 10, hence 
proportional to $E_4 E_6$. Due to the relation~(\ref{eq:BPSsum}) the space-time anomaly cancellation condition 
$n_H - n_V=244$ fixes the coefficient to one. Hence the expression~(\ref{eq:newk3t2}), obtained 
from the standard embedding with $(4,4)$ worldsheet supersymmetry, extends to any $(0,4)$ compactification; it means 
in particular that the "Mathieu moonshine" is a property of the $K3\times T^2$ new supersymmetric index regardless of the choice of 
gauge bundle~\cite{2013JHEP...09..030C}. Determining whether this property extends to Fu-Yau compactifications is one 
of the motivations for the present work.

\subsection{New supersymmetric index of Fu-Yau compactifications}

We now consider the main topic of this work, the computation of the new supersymmetric index of Fu-Yau compactifications based 
on their worldsheet formulation as torsion gauged linear sigma-models. 

The starting point of the computation is the same as for $\text{K3}\times T^2$ compactifications. However in the case of torsion GLSMs one cannot split the 
worldsheet theory as a tensor product of the $T^2$ and the K3 factors, as none of them makes sense as a quantum theory in isolation.  
We assume as before that the gauge bundle over the total space, which is the pullback of a Hermitian-Yang-Mills gauge bundle $V$ 
over the K3 base, is embedded as $V\subset SO(2n) \subset E_8$ in the first $E_8$ factor. 
				
In analogy with the $\text{K3}\times T^2$ case, we decompose the new supersymmetric index in terms of a twining partition function 
for the torsion GLSM as follows,
\begin{equation}
Z_\textsc{new}(\tau,\bar\tau)=\frac{\bar{\eta}^{2}E_{4}(\tau,0)}{2\eta^{10}}\sum_{\gamma,\delta=0}^{1}q^{\gamma^{2}}
\left.\left\{\left(\frac{\vartheta_{1}\left(\tau|y\right)}{\eta(\tau)}\right)^{8-n}
Z_\textsc{fy}\left(\tau,\bar{\tau},y\right)\right\}\right|_{y=\frac{\gamma\tau+\delta}{2}}\, ,\label{eq:index}
\end{equation}
where we have defined
\begin{equation}
Z_\textsc{fy}\left(\tau,\bar{\tau},y\right) = \frac{1}{\bar{\eta}(\bar{\tau})^{2}}
\text{Tr}_{\textsc{rr},\mathcal{H}_\textsc{fy}}\left[  e^{2i\pi y J_{0}}\bar{J}_{\, 0}
(-1)^{F} q^{L_0-c/24}\bar{q}^{\bar{L}_0-\bar{c}/24} \right]\, ,\label{eq:FYpart}
\end{equation}		
the trace being taken into the Hilbert space of the $(0,2)$ superconformal theory 
obtained as the infrared fixed point of the torsion GLSM.
	
A crucial point at this stage is that the right-moving fermions $(\chi,\bar \chi)$
associated with the $T^2$ factor, that belong to the torsion multiplet, are free in the 
Wess-Zumino gauge, see the Lagrangian~(\ref{eq:chiralcoupling}), in particular not coupled to the 
components of gauge multiplet; this is 
the feature of the theory that eventually leads to $\mathcal{N}=2$ supersymmetry in space-time.  
The right-moving R-current of the superconformal algebra, 
whose zero-mode $\bar{J}^0$ appears in the trace~(\ref{eq:FYpart}), is of the form 
\begin{equation}
\bar{J} = \bar{\chi} \chi +  \cdots \, ,
\end{equation}
where the ellipsis stands  for $(i)$ a term in $\bar \partial \alpha$, as the bottom 
component of the torsion multiplet can have a shift R-charge, $(ii)$ the contributions of the chiral and Fermi multiplets and 
$(iii)$ $\mathcal{Q}$-exact terms, where $\mathcal{Q}$ is the localization 
supercharge (see next section), relating the exact R-current to the Noether one defined in the UV theory. 
Because there are two  right-moving fermionic zero-modes $\chi_{0}$ and $\bar{\chi}_{0}$ that need to be saturated 
in the path integral, and that there are no interactions involving these fermionic fields in the Lagrangian,  we do not have to care about 
these extra terms in any case, as their contribution to the path integral vanishes.

In summary, the new supersymmetric index of Fu-Yau compactifications follows from the twisted partition function~(\ref{eq:FYpart}) that 
can be formulated as a path integral. Considering the theory on a two-dimensional Euclidean torus of complex structure $\tau$, 
the quantity to compute can be schematically written as 
\begin{multline}
\label{eq:FYcompact}
Z_\textsc{fy} (\tau,\bar \tau, y) = \frac{1}{\bar{\eta}(\bar{\tau})^{2}}\int \mathscr{D} a_{z} \mathscr{D} a_{\bar z}\mathscr{D} \lambda \mathscr{D} 
\bar{\lambda} \mathscr{D} D \, 
e^{-\frac{1}{e^2} S_{\text{v.m.}}[a, \lambda, D]- t\, S_{\textsc{fi}} (a,D)}\ \times \\ \times \ 
\int \mathscr{D} \phi_{i}\mathscr{D} \bar{\phi}_{i}  \mathscr{D} \psi_{i} \mathscr{D} \bar{\psi}_{i}  \, 
e^{-\frac{1}{g^2} S_{\text{c.m.}}[\phi_i,\psi_i,a,D,a_\textsc{l}]}
\ \times \\ \times \ 
\int \mathscr{D} \gamma_{a}\mathscr{D} \bar{\gamma}_{a}  \mathscr{D} G_a \mathscr{D} \bar{G}_{a}  \, 
e^{-\frac{1}{f^2} 
S_{\text{f.m.}}[\gamma_{a},G_a,a,a_\textsc{l}]-S_{\textsc{j}}[\gamma_{a},G_a,\phi_i,\psi_{i}]}
\ \times \\ \times \ 
\int \mathscr{D} \alpha \mathscr{D} \bar{\alpha}  \mathscr{D} \chi \mathscr{D} \bar{\chi}  \, 
e^{-S_{\text{t.m.}}[\alpha,\chi,a,a_\textsc{l}]}\int \frac{\d^2 z}{2\tau_2} \,  
\bar{\chi} \chi\, ,
\end{multline}
where we have included a background gauge field for the $U(1)_{\textsc{l}}$ global symmetry
\begin{equation}
\label{eq:backgd}
a_\textsc{l} = \frac{\pi y}{2i \tau_2} {\rm d}z-\frac{\pi y}{2i \tau_2} {\rm d}\bar z\, ,
\end{equation}
in order to implement the twisted boundary conditions.\footnote{Strictly speaking, we define the twisted path integral 
for real $y$ (corresponding to twisted boundary conditions along the space-like cycle) and consider an analytic 
continuation of the result, see~\cite{Murthy:2013mya}.} The torsion multiplet will be coupled chirally to this 
flat connection, in the same way as it couples to the dynamical gauge field, see eq.~(\ref{eq:chiralcoupling}).

The left- and right-moving fermions have periodic boundary conditions 
along both one-cycles of the worldsheet torus. We have also included for latter convenience coupling constants $1/g^2$ and 
$1/f^2$ in front of the chiral and Fermi multiplets actions, respectively $S_{\text{c.m.}}$ and $S_{\text{f.m.}}$, besides the usual $1/e^2$ 
factor in front of the vector multiplet action $S_{\text{v.m.}}$ and $t$ in front of the Fayet-Iliopoulos term  
$S_{\textsc{fi}}$. Finally $S_{\text{t.m.}}$ denotes the torsion multiplet action. 

To take care of the gauge redundancy one should in principle introduce 
a gauge-fixing procedure and the corresponding Faddeev-Popov ghosts; however it does not really impact the computation 
of the path integral through supersymmetric localization that will follow, see~\cite{Ashok:2013pya} for details. 

Having set the calculation in functional language will allow us to deal with it using localization techniques. In this formulation one 
sees that the insertion of the $\bar{J}_{0}$ operator only contributes through the free right-moving fermion $\chi$ which is part 
of the torsion multiplet, and this insertion appears as a prescription to deal with the fermionic zero modes. This will be important in proving that the 
supersymmetric localization method is valid in this context, as we shall explain below. 
							
%%%%%%%%%%%%%%%%%%%%%%%%%%%%%%%%%%%%%%%%%%%%%%%%%%%%%%%%%%%%%%%%%%%%%%%%%%%%%%%%%%%%%%%%%%%%%%%%%%%%%%%%%%%%%%%%%%%%%%%%%%%%%%%%%%%%%%%%%%%%%%%%%%%%%%%%%%%%%%%%%%%%%%%%%%%%%%%%%%%%%%%%%%%%%%%%%%%%%%%%%%%%%%%%%%%%%%%%%%%%%%%%%%%%%%%%%%%%%%%%%%%%%%%%%%%%%%%%%%%%%%%%%%%%%%%%%%%%%%%%%%%%%%%%%%%%%%%%%%%%%%%%%%%%%%%%%%%%%%%%%%%%%%%%%%%%%%%%%%%%%%%%%%%%%%%%%%%%%%%%%%%%%%%%%%%%%%%%%%%%%%%%%%%%%%%%%%%%%%%%%%%%%%%%%%%%%%%%%%%%%%%%%%%%%%%%%%%%%%%%%%%%%%%%%%%%%%%%%%%%%%%%%%%%%%%%%%%%%%%%%%%%%%%%%%%%%%%%%%%%%%%%%%%%%%%%%%%%%%%%%%%%%%%%%%%%%%%%%%%%%%%%%%%%%%%%%%%%%%%%%%%%%%%%%%%%%%%%%%%%%%%%%%%%%%%%%%%%%%%%%%%%%%%%%%%%%%%%%%%%%%%%%%%%%%%%%%%%%%%%%%%%%%%%%%%%%%%%%%%%%%%%%%%%%%%%%%%%%%%%%%%%
		
\section{New supersymmetric index through localization}
\label{sec:local}

In this section we obtain the twining partition function of Fu-Yau compactifications, defined by eq.~(\ref{eq:FYpart}), allowing 
to compute their new supersymmetric index using eq.~(\ref{eq:index}). In this section we consider  the case of a $U(1)$ 
worldsheet gauge group; the main result  is given by equation~(\ref{eq:oneloopdetrankone}). The generalization to higher rank 
will be provided in the next section. 

\subsection{Justification of the supersymmetric localization method}
				
Supersymmetric localization techniques  have been successfully applied to compute the elliptic genera of ordinary (0,2) gauged linear sigma-models, 
see~\cite{Gadde:2013dda,2014LMaPh.104..465B,2015CMaPh.333.1241B}. Our goal is to extend these results to the new supersymmetric index of Fu-Yau 
compactifications using the torsion gauged linear sigma-models. 

One immediate objection to this project is that, as mentioned above, the contribution of the torsion multiplet to the action is 
not invariant under the supersymmetry transformations~(\ref{eq:susytransfo}); furthermore, the operator insertion 
$\int\frac{\d[2]z}{2\tau_{2}}\ \bar{\chi}\chi$ in the path integral~(\ref{eq:FYcompact}) is  not supersymmetric. 
As we will see below, these two obstacles can be successfully overcome.  
						
The GLSM corresponding to the base contains a $(0,2)$ vector multiplet together with $(0,2)$ chiral and Fermi multiplets 
(conventions related to $(0,2)$ superspace are gathered in \cref{appendixConventions}). Define the supercharge, cf. equation~(\ref{eq:susytransfo}):
\begin{equation}
	\mathcal{Q}=\left(\delta_{\epsilon}\right)|_{\epsilon=\bar{\epsilon}=1}.
\end{equation}
As was noticed in~\cite{2014LMaPh.104..465B}, the Lagrangian describing the dynamics of these multiplets, including 
the superpotential term $\mathcal{L}_{\textsc{j}}$ and the Fayet-Iliopoulos term $\mathcal{L}_{\textsc{fi}}$, are actually exact 
with respect to the transformation introduced above. One finds
\begin{align}
\mathcal{L}_{\text{c.m.}}&=\mathcal{Q}\left(-2\bar{\phi}\nabla_{z}\psi+Q\bar{\phi}\bar{\lambda}\phi\right),
\notag\\
\mathcal{L}_{\text{f.m.}}&=\mathcal{Q}\left(\gamma\bar{G}\right) -\mathcal{Q}\left(\gamma J(\phi)\right),\notag \\
\mathcal{L}_{\text{v.m.}}&=\frac{1}{2}\mathcal{Q}\left(\left(a_{z\bar{z}}-D\right)\lambda\right) ,\\ 
\mathcal{L}_{\textsc{j}} & = -\mathcal{Q}\left(\gamma J(\phi)\right), \notag\\
\mathcal{L}_{\textsc{fi}} &=   \mathcal{Q}   \lambda \, . \notag
\end{align}
In an ordinary GLSM, this would imply immediately that the path integral is independent of the coupling constants $e$, $f$ and $g$ 
and of the FI parameter $t$.

To understand what happens in the present situation, let us write  the contribution of the base and of the vector multiplet to the TGLSM as 
$S_{\text{K3}}=\frac{1}{e^2} \mathcal{Q}\mu_{\text{v.m.}} +\mathcal{Q}\nu$ 
where the first term is the vector multiplet action, written as a $\mathcal{Q}$-exact term, and $\mathcal{Q}\nu$ denotes 
the ($\mathcal{Q}$-exact as well) contribution of the chiral and Fermi multiplets and of the constant FI term. The functional integral we aim 
to compute is of the schematic form 
\begin{equation}
Z_\textsc{fy} (\tau,\bar \tau, y)=\frac{1}{\bar{\eta}(\bar{\tau})^{2}}\int\D \Phi\D \Lambda\D A \mathscr{D} \Theta \ e^{-\frac{1}{e^2} 
\mathcal{Q}\mu_{\text{v.m.}} -\mathcal{Q}\nu}
e^{-S_{\text{t.m.}}[\Theta,A]}\int\frac{\d[2]z}{2\tau_{2}}\ \bar{\chi}\chi
\end{equation}
One considers then the derivative with respect to  $1/e^2$:
\begin{equation}
\frac{\partial Z_\textsc{fy} (\tau,\bar \tau, y)}{\partial (1/e^2)}=-\frac{1}{\bar{\eta}(\bar{\tau})^{2}}\int\mathscr{D} \Phi\mathscr{D} \Lambda\mathscr{D}
A \mathscr{D} \Theta \ \mathcal{Q} \mu_{\text{v.m.}}\ e^{-\frac{1}{e^2} \mathcal{Q}\mu_{\text{v.m.}} -\mathcal{Q}\nu}
e^{-S_{\text{t.m.}}[\Theta,A]}\int\frac{\d[2]z}{2\tau_{2}}\ \bar{\chi}\chi \, .
\end{equation}

As mentioned above, the operator insertion $\int\frac{\d[2]z}{2\tau_{2}}\ \bar{\chi}\chi$ has the effect of saturating the fermionic zero modes present 
in the measure $\mathscr{D} \Theta$ over the torsion multiplet. Hence its variation under the action of the supercharge $\mathcal{Q}$, while non-zero,  leads to terms 
which do not saturate the fermionic zero modes anymore, and thus do not contribute to the path integral. 

Since the supersymmetry transformation we are considering contains a supergauge transformation of chiral parameter $ \Xi_{wz } |_{\bar{\epsilon}=1}$, 
see eq.~(\ref{eq:susytransfo}), there is a non-trivial transformation of the functional measure over the chiral and Fermi multiplets due to the gauge 
anomaly. At the same time, the torsion multiplet action is not classically invariant under the action of the supercharge, see eq.~(\ref{eq:deltator}). 
Whenever the tadpole condition~(\ref{eq:tadpole}) is satisfied, these two variations cancel each other:
\begin{equation}
\mathcal{Q} \left(  \mathscr{D} \Phi\mathscr{D} \Lambda e^{-S_{\text{t.m.}}[\Theta,A]} \right)=0\, .
\end{equation}

In conclusion, whenever the quantum anomaly of the base GLSM is canceled against the classical contribution 
from the torsion multiplet, we get as in more familiar examples
\begin{equation}
\frac{\partial}{\partial (1/e^2)}Z_\textsc{fy} (\tau,\bar \tau, y) = -\frac{1}{\bar{\eta}(\bar{\tau})^{2}}
\int \mathcal{Q}\left(\mathscr{D} \Phi \mathscr{D} \Lambda \mathscr{D}  A\  
e^{-\frac{1}{e^2} \mathcal{Q}\mu_{\text{v.m.}} -\mathcal{Q}\nu-W[A]}\, \mu_{\text{v.m.}} \int\frac{\d[2]z}{2\tau_{2}}\ \bar{\chi}\chi\right) = 0 \, ,
\end{equation}
using an analogue of Stokes' theorem in field space. The result of the path integral is then independent of the gauge coupling, allowing to take a free-field limit $e\to 0$. The same 
reasoning allows to take the limit $g\to 0$ and $f\to 0$ in the chiral and Fermi multiplets actions respectively. By rescaling the superfields $\Phi' = \Phi/g$ and 
$\Gamma ' = \Gamma /f$ one sees that the superpotential couplings do not contribute to the path integral which is localized in the free-field limit of the theory, 
as far as the base  GLSM is concerned.

A similar argument regarding the dependence of the path integral on the torsion multiplet couplings would fail, as the torsion multiplet 
action is not $\mathcal{Q}$-exact, being not even $\mathcal{Q}$-closed.\footnote{Even if it were the case (this is what happens if 
one considers $K3\times T^2$ examples), the $\mathcal{Q}$ variation of $\int\frac{\d[2]z}{2\tau_{2}}\ \bar{\chi}\chi$ would then contribute 
to the path integral.} Nevertheless, this action is Gaussian hence the path integral can be performed exactly. As expected, it implies that the result of the path 
integral computation does depend on the moduli $(T,U)$ of the principal two-torus bundle in the Fu-Yau geometry.

As in~\cite{2014LMaPh.104..465B} the localization locus contains the following zero-modes, that should be integrated over:
\begin{itemize}
\item Gauge holonomies on the worldsheet two-torus, parametrized by $a = \frac{\pi\bar u}{2i \tau_2} {\rm d}z-\frac{\pi u}{2i \tau_2} {\rm d}\bar z$, 
$u$ being defined on the torus of complex structure $\tau$ to avoid gauge redundancy,
\item The zero-mode $D_0$ of the auxiliary field in the gauge multiplet,
\item The gaugino zero-modes $\lambda_0$, $\bar{\lambda}_0$. 
\end{itemize}

Setting aside the contribution of the torsion multiplet, most of the steps that go into the derivation of the elliptic genus by Benini, Eager, Hori and 
Tachikawa~\cite{2014LMaPh.104..465B}, especially the reduction of the integral over the gauge holonomies into a contour 
integral of the one-loop determinants, carry over to the present situation without significant modifications. 
We refer the reader to this article for a detailed account of the computation and provide below justifications of this statement.  

In order to saturate the gaugino zero-modes, the contribution from the chiral multiplets at one-loop is of the form, 
in the limit $e\to 0$, 
\begin{equation}
f_{\text{c.m.}} (\tau, y, u, D_0)=
\int\d\lambda_{0}\d\bar{\lambda}_{0}
\left\langle\left|\int\d[2]z\sum_{i}Q_{i}\lambda\bar{\psi}_{i}\phi_{i}\right|^{2}\right\rangle_{\text{free}}\, .
\end{equation}
Since the torsion multiplet has no coupling to the gaugini, as was explained in section~\ref{sec:GLSM}, it is not involved in the 
saturation of their zero-modes; hence this part of the derivation is unchanged. Furthermore, the torsion multiplet has no coupling to 
the auxiliary $D$-field (by supersymmetry); as a consequence, the $\bar u$-dependence of the one-loop determinant lies entirely in the contribution 
from the chiral multiplets of the base, allowing to reduce the integral over the $u$-plane to a contour integral as 
in~\cite{2014LMaPh.104..465B}.\footnote{This will be true only if the quantum anomaly from the base GLSM is canceled 
by the gauge non-invariance of the torsion multiplet, see eq.~(\ref{eq:oneloopdet}) below.} Then, the singularities that arise in the limit $e\to 0$ 
after integrating over the zero-mode $D_0$, as well as the contour deformation in the 
$D_0$-plane leading to the contour prescription,  remain the same as in the aforementioned computation. 

For a rank-one gauge group, the formula for the twining partition function~(\ref{eq:FYpart}) is then of the form:
\begin{equation}
Z_\textsc{fy} (\tau,\bar \tau, y)=\pm\frac{1}{2i\pi}\sum_{u^{\star}
\in\mathcal{M}_{\text{sing}}^{\pm}}\oint_{\mathcal{C}(u^{\star})}\ \Sigma_{\text{1-loop}}(\tau,\bar{\tau},y,u)\, ,
\end{equation}
where 
\begin{equation}
\Sigma_{\text{1-loop}}(\tau,\bar{\tau},y,u)=\frac{1}{\bar{\eta}(\bar{\tau})^{2}}Z_{A}\times
\left(\prod_{\Phi_{i}}Z_{\Phi_{i}}\right)\times
\left(\prod_{\Gamma_{a}}Z_{\Gamma_{a}}\right)\times Z_{\chi}\times Z_{\text{torus}}\, ,\label{genus}
\end{equation}
the various factors in the above formula being the one-loop contributions of the various multiplets around the localization locus, and $\mathcal{C}(u^{\star})$  
denoting a contour around the singularity $u^{\star}$. $\mathcal{M}_{\text{sing}}^{+}$ and $\mathcal{M}_{\text{sing}}^{-}$ form a 
partition of the set of poles of the product of chiral multiplet determinants and are described in detail 
in~\cite{2014LMaPh.104..465B}. These poles correspond to 'accidental' bosonic zero-modes and occur whenever
\begin{equation}
Q_i u + q^\textsc{l}_i y = 0 \mod \mathbb{Z} + \tau \mathbb{Z}\, ,
\end{equation}
$Q_i$ (resp $q^\textsc{L}_i$) being the gauge (resp $U(1)_{\textsc{l}}$) charge of the multiplet. The set of poles $\mathcal{M}_{\text{sing}}$ is 
then split into two sets according to $Q_i>0$ or $Q_i<0$.  Notice that the choice of $\mathcal{M}_{\text{sing}}^{+}$ or $\mathcal{M}_{\text{sing}}^{-}$ 
in~(\ref{genus}) give the same result since the sum of the residues of a meromorphic function on the torus vanishes. 

In the case where the gauge group $G$ has an arbitrary rank, the formula generalizes using a notion of residue in higher dimensions, 
the Jeffrey-Kirwan residue~\cite{Jeffrey1995291}; one obtains the following expression for the 
twining partition function in terms of the one-loop determinant~\cite{2015CMaPh.333.1241B}:
\begin{equation}
Z_\textsc{fy} (\tau,\bar \tau, y)=\frac{1}{|W|}\sum_{u^{\star}\in\mathcal{M}_{\text{sing}}}\underset{u=u^{\star}}{\text{JK-Res}}
\left(\text{Q}(u^{\star}),\eta\right)\Sigma_{\text{1-loop}}\, ,
\end{equation}
with now $\Sigma_{\text{1-loop}}$ a meromorphic $\text{rank}(G)$-form, and $|W|$ the order of the Weyl group. The sum does not depend on the 
choice of co-vector $\eta\in\mathfrak{h}^{*}$, in the dual of the Cartan subalgebra, per singular locus $u^{\star}$.

\subsection{\texorpdfstring{Contribution of the \text{K3}}{K3} base}
			
As we have noticed previously, the contributions from the chiral and Fermi multiplets corresponding to the $K3$ base, as well 
as from the gauge multiplets, are similar to those appearing in the elliptic genus computed 
in~\cite{2014LMaPh.104..465B}. However, since in the present context issues of gauge invariance are crucial, 
we need to be a little bit more careful regarding the definition of the chiral fermionic determinants. 
In the end, taking into account the contribution of the torsion 
multiplet, the tadpole condition will translate into a cancellation of the prefactors in these expressions.
				
In order to define the determinant of a chiral Dirac operator $\nabla (u)$ coupled to a (background) flat gauge field,  
one has to specify a way to split the determinant of the self-adjoint operator $\nabla^{\dagger}(u)\nabla(u)$ into a 'holomorphic' part and 
an 'anti-holomorphic' part. According to Quillen's theorem~\cite{Quillen}, the zeta-regularized determinant of the former 
is given by (see $e.g.$~\cite{alvarez-gaume1986} for a discussion in a similar context):
\begin{equation}
\text{Det}_\zeta\, \nabla(u)^{\dagger}\nabla(u)=e^{\frac{\pi}{\tau_{2}}(u-\bar{u})^{2}}|\vartheta_{1}(\tau|u)|^{2}\, ,
\end{equation}
where $u$ is here a compact notation which takes into account both the $U(1)$ gauge field and the background $U(1)_{\textsc{l}}$. 
Splitting $(u-\bar{u})^{2}=(u^{2}-u\bar{u})+(\bar{u}^{2}-u\bar{u})$, one can define the chiral determinant as:
\begin{equation}
\label{eq:chiraldet}
\text{Det}\, \nabla(u)=e^{\frac{\pi}{\tau_{2}}(u^{2}-u\bar{u})}\vartheta_{1}(\tau|u)\, ,
\end{equation}
modulo an overall factor independent of $u$; other definitions can be interpreted as corresponding 
to different choices of local counterterms. 

With this prescription, as was argued by Witten in~\cite{1997JGP....22..103W} in a related context, 
the gauge functional obtained after the path integral over the fermionic degrees of freedom can be 
viewed as a holomorphic section of a holomorphic line bundle over the space of gauge connections.  The determinant is indeed  
annihilated by the covariant derivative \mbox{$\frac{D}{D\bar{u}}=\frac{\partial}{\partial\bar{u}}+\frac{\pi}{\tau_{2}}u$} 
(restricted to its zero-mode part in the present situation). It turns out that this choice, besides its nice geometrical interpretation, is 
naturally compatible with the contribution from the torsion multiplet Lagrangian, see~\cref{eq:rayon_k/l} below, leading to an expression 
without modular anomalies. 
					
Equipped with this result, one can express the contribution of a $(0,2)$ chiral multiplet $\Phi_{i}$ of 
gauge charge $Q_{i}$ and $U(1)_{\textsc{l}}$ charge $q^\textsc{l}_{i}$, and a Fermi multiplet $\Gamma_{a}$ of gauge charge $Q_{a}$ and 
$U(1)_{\textsc{l}}$ charge $q^\textsc{L}_{a}$ as:
\begin{subequations}
\begin{align}
Z_{\Phi_{i}}(\tau,u,y)&=ie^{-\frac{\pi}{\tau_{2}}(\upsilon^{2}-\upsilon\bar{\upsilon})}\frac{\eta(\tau)}{\vartheta_{1}(\tau|\upsilon)}
& , \  \upsilon = Q_i u + q^\textsc{l}_{i}y
\, ,\\
Z_{\Gamma_{a}} (\tau,u,y) &=ie^{\frac{\pi}{\tau_{2}}(\upsilon^{2}-\upsilon\bar{\upsilon})}\frac{\vartheta_{1}(\tau|\upsilon)}{\eta(\tau)}
&  , \ \upsilon = Q_a u + q^\textsc{l}_{a}y \, .
\end{align}
\end{subequations}

Finally, the contribution from the vector multiplet reads, considering a $U(1)$ gauge group,
\begin{equation}
Z_{A} (\tau,y) =-2i\pi\eta(\tau)^{2} \d u \, .
\end{equation}
The final step will be to evaluate the contribution of the torsion multiplet. 			
							
\subsection{Torsion multiplet determinant}

In this section we derive the contribution of the torsion multiplet to the partition function $Z_\textsc{fy} (\tau,\bar \tau, y)$ at the localization locus. 
In the functional integral formulation it takes the form
\begin{equation}
\int\D \alpha_{1}\D \alpha_{2}\D \chi\D \bar{\chi}\ e^{-S_{\text{torsion}}[\alpha_{1},\alpha_{2},\chi,\bar{\chi},a,a_{\textsc{l}}]}\int\frac{\d[2]z}{2\tau_{2}}\ \bar{\chi}\chi.
\end{equation}
One is thus dealing with two compact bosons chirally coupled to a flat gauge field, together with a free right-moving Weyl fermion. As was 
noticed before, the action is not $\mathcal{Q}$-exact, however it is Gaussian hence can be computed explicitly. 

Let us evaluate first the contribution of the free fermion, which is completely decoupled from the vector multiplet. 
Taking into account the insertion $\int\frac{\d[2]z}{2\tau_{2}}\ \bar{\chi}\chi$ into 
the functional integral,\footnote{In our conventions, $\int\d[2]z=2\tau_{2}$} one gets, using formula~(\ref{propTheta}):
\begin{equation}
Z_{\chi}(\bar \tau) = \partial_\mu 
\int\D \chi\D \bar{\chi}\ e^{-\int\d[2]z\ 2\bar{\chi}\partial\chi+\mu\int\frac{\d[2]z}{2\tau_{2}}\ \bar{\chi}\chi}
\Big|_{\mu = 0} = \bar{\eta}(\bar{\tau})^{2}\, .
\end{equation}

\subsubsection{Orthogonal torus}
	
Now we compute the contribution from the 'axion field' $\alpha=\alpha_{1}+T\alpha_{2}$. It corresponds to  a pair  
of chiral bosons coupled to a gauge field $(a_z,a_{\bar{z}})$. Considering first an orthogonal torus with  no  B-field, 
for each of them one has to compute a path integral of the form 
\begin{equation}
\label{eq:chirallag}
\int\D \varphi\ \exp\left\{-\frac{R^{2}}{2\pi}\int\d[2]z\, 
\left( \partial\varphi\bar{\partial}\varphi+2a_{\bar{z}}\partial\varphi+a_{z}a_{\bar{z}}\right) \right\}\, ,
\end{equation}
where $\varphi \sim \varphi + 2\pi$. Here $a_{z}$ and $a_{\bar{z}}$ refer to both the dynamical gauge field and to the 
background $U(1)_{\textsc{l}}$ gauge field. In the present context, because of localization one focuses on the zero mode of the 
gauge fields, i.e. the holonomies on the worldsheet two-torus. 

At the fermionic radius  $R_{\text{f}}=\frac{1}{\sqrt{2}}$, the bosonic action appearing in the path integral~(\ref{eq:chirallag}) is actually nothing but the 
bosonized form of the chiral determinant~(\ref{eq:chiraldet}) (up to an anti-holomorphic determinant independent of the gauge field) that was considered by Witten 
in~\cite{1997JGP....22..103W}, hence motivating the choice made above for the latter. In the bosonic formulation this prescription amounts to set the coefficient 
of the $a_{z}a_{\bar{z}}$ term, which can be affected by local counterterms, to one, and implies that the classical variation of the Lagrangian 
under a gauge transformation is independent of $\varphi$.

The different instanton sectors of the free compact boson $\varphi$ on the torus obey the periodicity condition
\begin{equation}
\varphi\left(z+k+\tau l,\bar{z}+k+\bar{\tau}l\right)=\varphi\left(z,\bar{z}\right)+2\pi\left(km+ln\right)\, ,
\end{equation}
with winding numbers $m,n\in\mathbb{Z}$. The solution for the zero-modes $\phi_0$ is then given by: 
\begin{equation}
\varphi_{0}\left(z,\bar{z}\right)=\frac{i\pi}{\tau_{2}}\left(z\left(m\bar{\tau}-n\right)-\bar{z}\left(m\tau-n\right)\right)\, .
\end{equation}
Plugging this into the classical action and adding the contribution from the quantum fluctuations leads to 
\begin{multline}
Z_{S^{1}}(\tau,\bar \tau,u,y)= \exp\left(-\frac{2\pi}{\tau_{2}}R^{2}u\bar{u}\right) \frac{R}{\sqrt{\tau_{2}}|\eta(\tau)|^{2}}
\times \\ \times\ 
\sum_{m,n\in\mathbb{Z}} 
\exp\left(-\frac{\pi R^{2}}{\tau_{2}}|m\tau-n|^{2}-2iR^{2}\left(m\bar{\tau}-n\right)\left(a_{\bar{z}}\right)_{0}\right)\, .
\end{multline}
	
Poisson resummation formula (\cref{appendixTheta}) applied to the dummy variable $n$ allows to rewrite this as
\begin{align}
\label{eq:partcomp}
Z_{S^{1}}(\tau,\bar \tau,u,y)&=\frac{1}{|\eta(\tau)|^{2}} 
\exp\left(-\frac{2\pi}{\tau_{2}}R^{2}(u\bar{u}-u^2)\right)\times\nonumber\\
&\ \ \times\sum_{m,n\in\mathbb{Z}}\exp\left(\frac{i\pi\tau}{2}\left(\frac{n}{R}+Rm\right)^{2}
-\frac{i\pi\bar{\tau}}{2}\left(\frac{n}{R}-Rm\right)^{2}-i\pi Ru\left(\frac{n}{R}+Rm\right)\right)\, ,
\end{align}
Setting the radius to the free fermion radius $R_{\text{f}}=\frac{1}{\sqrt{2}}$, the above expression can be recast as a 
finite sum over the different spin structures on the worldsheet torus, namely 
\begin{multline}
\label{eq:rayfer}
Z_{S^{1}}\left(\tau,\bar \tau,u,y;R_{\text{f}}\right)=\frac{1}{2}\frac{1}{|\eta(\tau)|^{2}}
e^{-\frac{2\pi}{\tau_{2}}R_{\text{f}}^{2}\left[(Mu+m_{\textsc{l}}y)(M\bar u+m_{\textsc{l}}y)-(Mu+m_{\textsc{l}}y)^{2}\right]}\times \\ \times\ 
\sum_{k,l=0}^{1}\vartheta\left[^{k}_{l}\right](\tau|Mu+m_{\textsc{l}}y)\bar{\vartheta}\left[^{k}_{l}\right](\bar{\tau}|0)\, ,
\end{multline}
where we have restored the shift charge $M$ and added the coupling to the background $U(1)_{\textsc{l}}$ gauge field. Taking into account the second $S^{1}$ is 
straightforward, since the two circles factorize.

Comparing the holomorphic part of the partition function~(\ref{eq:rayfer}) to the contribution of a left-moving fermion coming 
from a charged Fermi multiplet of the base GLSM, one has in the former case an independent sum over the spin structures $(k,l)$ on the 
worldsheet two-torus, while in the latter case the spin structure is chosen periodic along both one-cycles.\footnote{We consider in this 
discussion that we are in the left GSO sector given by $y=0$ for simplicity. 
Considering a different sector does not change the outcome of the argument; the important point is 
that the spin structures of all free Fermi multiplets are identical (considering that the gauge group lies in a single $E_8$).} 
This simple observation clarifies some statements about topology-changing T-dualities, mixing 
the torus and gauge bundles, that were originally proposed by Evslin and Minasian in~\cite{Evslin:2008zm} in the effective 
theory context, and discussed by one of the authors in the torsion GLSM framework~\cite{2013JHEP...11..093I}  
(see also~\cite{2007hep.th....3048A} for related comments). Such duality, 
that exchanges  a line bundle over the base $\mathcal{S}$ and a circle bundle at the fermionic radius,  
is indeed a symmetry of the twining partition function $Z_\textsc{fy}$ built from~(\ref{eq:rayfer}) only in the sector $(k=0,l =0)$, in which case 
the two corresponding left-moving fermions, from the Fermi multiplet and from the left-moving component of the fermionized $S^1$ fiber, have 
identical (odd) spin structure. Including the independent sum over the spin structures $(k,l)$ of the latter does not respect this symmetry.

In order to generalize the results obtained at the fermionic radius to compact bosons of arbitrary rational radius squared, 
it is convenient to rewrite the previous expression in terms of $\widehat{\mathfrak{su}(2)}$ theta functions at level $2$ 
(see appendix~\ref{appendixTheta}):
\begin{multline}
Z_{S^{1}}\left(\tau,\bar \tau,u,y;R_{\text{f}}\right)=\frac{1}{|\eta(\tau)|^{2}}
e^{-\frac{2\pi}{\tau_{2}}R_{\text{f}}^{2}\left[(Mu+m_{\textsc{l}}y)(M\bar u+m_{\textsc{l}}y)-(Mu+m_{\textsc{l}}y)^{2}\right]}
\times \\ \times\ 
\sum_{s\in\mathbb{Z}_{4}}\Theta_{s,2}\left(\tau\left|2(Mu+m_{\textsc{l}}y)\right.\right)\bar{\Theta}_{s,2}(\bar{\tau}|0)\, ,\label{FermionRadius}
\end{multline}
although the sum over spin structures is no longer explicit. 

Whenever a compact boson is at radius $R=\sqrt{\frac{k}{l}}$ (with $k$ and $l$ coprime integers), the corresponding $c=1$ 
conformal field theory becomes rational. One can then reorganize the sum in~(\ref{eq:partcomp}) over 
infinitely many $\widehat{u(1)}_L \times \widehat{u(1)}_R$ representations into a finite sum over representations of 
the chiral algebra, much as in the case of the fermionic radius that we have discussed previously. 
In terms of $\widehat{\mathfrak{su}(2)}$ theta functions level $kl$, one obtains then 
\begin{multline}
Z_{S^{1}}\left(\tau,\bar \tau,u,y;R=\sqrt{\tfrac{k}{l}}\right)=\frac{1}{|\eta(\tau)|^{2}}
e^{-\frac{2\pi}{\tau_{2}}R^{2}\left[(Mu+m_{\textsc{l}}y)(M\bar u+m_{\textsc{l}}y)-(Mu+m_{\textsc{l}}y)^{2}\right]}
\times \\
\times\sum\limits_{s,\bar s \in \mathbb{Z}_{2kl},\, \left\{ \substack{s+\bar{s}=0\ \ [2k]\\s-\bar{s}=0\ \ [2l]}\right.}
\Theta_{s,kl}\left(\tau\left|\tfrac{2}{l}(Mu+m_{\textsc{l}}y)\right.\right)\bar{\Theta}_{\bar{s},kl}(\bar{\tau}|0)\, .\label{eq:rayon_k/l}
\end{multline}
				
\subsubsection{Arbitrary rational torus}
	
As we have reviewed in section~\ref{sec:GLSM}, covariance under $O(2,2;\mathbb{Z})$ implies that 
the moduli of the (spacetime) two-torus should always be those of a rational $c=2$ conformal 
field theory, $i.e.$ with $T$ and $U$ belonging to the same imaginary 
quadratic number field $\mathbb{Q}(\sqrt{D})$ with $D<0$. Reducing the corresponding Narain lattice 
to a sum over characters of the underlying chiral algebra of the rational theory can be done 
explicitly, as previously, for any given example. However to write down the result explicitely in 
a uniform way for all cases requires a more abstract point of view. 

A precise relation between rational Narain lattices and the data of rational conformal field theories with $c=2$ was given 
in a beautiful article by Hosono, Lian, Oguiso and Yau~\cite{2003CMaPh.241..245H}.  Whenever the  Narain lattice $\Gamma^{2,2} (T,U)$ is rational, 
$i.e.$ with $T$ and $U$ belonging to the same $\mathbb{Q}(\sqrt{D})$, the even positive definite lattices
\begin{equation}
\Pi_\textsc{l} : =  \Gamma^{2,2} (T,U) \cap \mathbb{R}^{2,0} \ , \quad \Pi_\textsc{r} : =  \Gamma^{2,2} (T,U) \cap \mathbb{R}^{0,2}
\end{equation}
have rank two. Conversely, a rational CFT with $c=2$ is given by a {\it triple} $(\Gamma_{\textsc{l}},\, \Gamma_{\textsc{r}}, \, \phi)$, where $\Gamma_{\textsc{l,r}}$ 
are even positive definite lattices of rank two, and $\phi$ an isometry mapping one discriminant group to 
the other, $i.e.$ an application $\phi: \Gamma_{\textsc{l}}^{\vee}/\Gamma_{\textsc{l}} 
\to \Gamma_{\textsc{r}}^{\vee}/\Gamma_{\textsc{r}}$, with $\Gamma_{\textsc{l}}^{\vee}=\text{Hom}(\Gamma_{\textsc{l}},\mathbb{Z})$, 
preserving the bilinear form; it is known also as the gluing map. 

To each even positive definite lattice of rank two one can associate an integral quadratic form $Q= ax^2 + bxy + cy^2$ with, choosing 
a basis, $2a=(e_1,e_1)$, $b=(e_1,e_2)$ and  $2c=(e_2,e_2)$. The $GL_2 (\mathbb{Z})$ equivalent classes of quadratic forms, $\bar{\cal{C}}$, 
are isomorphic to the $GL_2 (\mathbb{Z})$ equivalent classes  of even positive definite lattices of rank two, $[\Gamma ]$, 
characterized by their invariant discriminant (resp. determinant) $4ac-b^2 : = -D$. Restricting the former classes to $SL_2 (\mathbb{Z})$ equivalence 
classes $\mathcal{C}$, one obtains for each $D$ an Abelian group of finite rank, equipped with a composition law 
known as the {\it Gauss product} $\mathcal{C}\star\mathcal{C}'$.\footnote{Whenever the quadratic forms are not primitive, $i.e.$ 
such that $\text{gcd}\, (a,b,c)>1$, these statements should be slightly modified, see~\cite{2003CMaPh.241..245H} for details.}

Likewise, for a given determinant $D$, the equivalence classes of rational Narain lattices $ \Gamma^{2,2} (T,U)$ 
under the $SL_2 (\mathbb{Z})\times SL_2 (\mathbb{Z})$ action on  $T$ and $U$ correspond 
to equivalent classes of quadratic forms, $\mathcal{C}_T$ and $\mathcal{C}_U$. 
The equivalence classes of the left and right lattices defining the rational CFT, 
$[\Gamma_{\textsc{l}}]$ and $[\Gamma_{\textsc{r}}]$, are then given in terms of the equivalent classes of the moduli  by:
\begin{equation}
[\Gamma_{\textsc{l}}] = \left[\Gamma_{\overline{\mathcal{C}_T \star (\mathcal{C}_U)^{-1}}} \right]\ , \quad 
[\Gamma_{\textsc{r}}] = \left[\Gamma_{\overline{\mathcal{C}_T \star \mathcal{C}_U}}\right] \, .
\end{equation}
This provides the data of the $c=2$ rational CFT for any rational $(2,2)$ Narain lattice. 

One can now express the result of the path integral in terms of the 
rational CFT data, in other words in terms of the theta-functions $\Theta^{\Gamma}_{\mu}$ associated with the left and 
right lattices $\Gamma_{\textsc{l}}$ and $\Gamma_{\textsc{r}}$, see~\cref{appendixTheta} for details. Let us consider 
the case of a $U(1)$ (worldsheet) gauge group. Taking an orthonormal basis, the 
two-dimensional vector of topological charges 
corresponding to the two-torus bundle, see section~\ref{sec:GLSM}, is of the form
\begin{equation}
p_\textsc{m}=\sqrt{\frac{2U_{2}}{T_{2}}}\begin{pmatrix}M_{1}+T_{1}M_{2}\\T_{2}M_{2}\end{pmatrix}\, .\label{vectorTheta}
\end{equation}
One can check that this vector belongs actually to the lattice $\Gamma^{2,2} (T,U) \cap \mathbb{R}^{2,0}$, hence to the left lattice 
of the rational CFT. A convenient assignment of $U(1)_{\textsc{l}}$ shift charges for cancellation of global anomalies is 
to take them proportional to the gauge charges, with a  coefficient of proportionality  $\lambda$, see section~\ref{sec:examples}.

One obtains then the following one-loop contribution from the bosonic degrees of freedom of the 
torsion multiplet, for arbitrary $T$ and $U$ in the same imaginary quadratic number field $\mathbb{Q}(\sqrt{D})$:
\begin{multline}
Z_{\text{torus}} (\tau,\bar \tau,u,y;T,U)= \exp \left(-\frac{2\pi}{\tau_2} 
\left[ (u+\lambda y)(\bar u+\lambda y) - (u+\lambda y)^2 \right] \langle p_\textsc{m},p_\textsc{m}\rangle_{\Gamma_\textsc{l}} 
\right) \times \\ \times \ 
\sum_{\mu\in\Gamma_{\textsc{l}}^{\vee}/\Gamma_{\textsc{l}}}
\frac{\Theta_{\mu}^{\Gamma_{\textsc{l}}}(\tau|(u+\lambda y)p_\textsc{m})}{\eta(\tau)^{2}}
\frac{\bar{\Theta}_{\varphi(\mu)}^{\Gamma_{R}}(\bar{\tau}|0)}{\bar{\eta}(\bar{\tau})^{2}}\, .
\label{eq:torusrat}
\end{multline}
where $\langle \star,\star \rangle_{\Gamma_\textsc{l}}$ is the inner product on $\Gamma_{\textsc{l}}$. Hence the quadratic 
prefactor of this expression is written in terms of the norm of the vector 
$p_\textsc{m} \in \Gamma_\textsc{l}$ giving the topological charges of the torus bundle
\begin{equation}
\label{eq:chargebundle}
\langle p_\textsc{m},p_\textsc{m}\rangle_{\Gamma_\textsc{l}}  = \frac{2U_2}{T_2} |M|^2\, .
\end{equation}
$O(2,2;\mathbb{Z})$ T-duality transformations are mapped, under the correspondence between rational Narain lattices and rational CFTs 
summarized above, to isometries of the triple providing the rational CFT data, hence preserve~(\ref{eq:chargebundle}). It provides 
an elegant explanation of the invariance of this expression, which gives also the contribution of the torus bundle to the integrated 
Bianchi identity, under the perturbative duality group~\cite{Evslin:2008zm}; as we have shown, this property is intimately related 
to the rational nature of the Narain lattice. 
						
\subsection{The result}
\label{subsec:result}

Having dealt separately with the contribution of each type of multiplet, one can write the full one-loop determinant, 
in the case of a rank one gauge group, 
as follows:
\begin{align}
\label{eq:oneloopdet}
\Sigma_{\text{1-loop}} (\tau, \bar \tau, y)
=-2i\pi\eta(\tau)^{2}&\prod_{\Phi_{i}}\frac{i\eta(\tau)}{\vartheta_{1}(\tau|Q_{i}u+q_{i}^{\textsc{l}}y)}
\, \prod_{\Gamma_{a}}\frac{i\vartheta_{1}(\tau|Q_{a}u+q^{\textsc{l}}_{a}y)}{\eta(\tau)}\ \times\nonumber\\
&\times\sum_{\mu\in\Gamma_{\textsc{l}}^{\vee}/
\Gamma_{\textsc{l}}}\frac{\Theta_{\mu}^{\Gamma_{\textsc{l}}}(\tau|(u+\lambda y)p_\textsc{m})}{
\eta(\tau)^{2}}\frac{\bar{\Theta}_{\varphi(\mu)}^{\Gamma_{R}}(\bar{\tau}|0)}{\bar{\eta}(\bar{\tau})^{2}}\ \times\nonumber\\
&\times\exp\Big[\underbrace{\Big(-\sum_{i}Q_{i}^{\ 2}+\sum_{a}Q_{a}^{\ 2}+\frac{2U_{2}}{T_{2}}|M|^{2}\Big)}_{=\, 0}
\left(u^{2}-u\bar{u}\right)\Big]\ \d u\, .
\end{align}
The first line corresponds to the K3 base and the second line to the two-torus fiber.

In this computation we had chosen a prescription for the determinant of a chiral Dirac operator consistent with the 
torsion multiplet contribution, such that at the end the various factors present in the last line of eq.~(\ref{eq:oneloopdet}) 
cancel each other whenever the gauge charges satisfy the tadpole condition~(\ref{eq:tadpole}). Notice that 
there are also factors linear in $y$, corresponding to the $U(1)_{\textsc{l}}$ global anomaly, that we did not include in the equation for 
sake of clarity; likewise, they cancel among themselves in an anomaly-free model.

Thus the one-loop determinant $\Sigma_{\text{1-loop}}$ is  a holomorphic function of the gauge field holonomy. Simplifying the above expression 
and injecting it in the contour integral, one gets for the twining partition function~(\ref{eq:FYpart}) the result:
\begin{align}
Z_\textsc{fy} (\tau,\bar \tau, y)=\pm\eta(\tau)^{2}\sum_{u^{\star}
\in\mathcal{M}_{\text{sing}}^{\pm}}\oint_{\mathcal{C}(u^{\star})}\d u\ \prod_{\Phi_{i}}&
\left(i\frac{\eta(\tau)}{\vartheta_{1}(\tau|Q_{i}u+q_{i}^{\textsc{l}}y)}\right)
\prod_{\Gamma_{a}}\left(i\frac{\vartheta_{1}(\tau|Q_{a}u+q^{\textsc{l}}_{a}y)}{\eta(\tau)}\right)\times\nonumber\\
&\times\, \sum_{\mu\in\Gamma_{\textsc{l}}^{\vee}/\Gamma_{\textsc{l}}}
\frac{\Theta_{\mu}^{\Gamma_{\textsc{l}}}(\tau|(u+\lambda y)p_\textsc{m})}{\eta(\tau)^{2}}
\frac{\bar{\Theta}_{\varphi(\mu)}^{\Gamma_{R}}(\bar{\tau}|0)}{\bar{\eta}(\bar{\tau})^{2}}\, .	
\label{eq:oneloopdetrankone}			
\end{align}
This quantity is then plugged into eq.~\eqref{eq:index} to finally give the new supersymmetric index of Fu-Yau compactifications.  

For consistency the new supersymmetric index that we have computed should behave 
properly under transformations of the modular group $PSL(2,\mathbb{Z})_\tau$. 
As mentioned above, $\tau_{2}Z_{\textsc{new}}(\tau,\bar{\tau})$ should be a (non-holomorphic) modular form of weight $-2$. 
Tracking this statement back to the modular behaviour of the non-holomorphic twining partition function $Z_{\textsc{fy}}$, 
one should check that whenever the anomaly cancellation condition is satisfied, 
the latter behaves as a weak Jacobi form of weight $0$ and index $\frac{r}{2}$, where $r$ is the rank of the vector bundle, 
although it is not holomorphic in $\tau$.
		
This behavior will be checked first at the level of the torsion GLSM in the next section once we have given the charge assignment,
and then later on using the geometrical formula that we will define is section \ref{sec:geom}.

%%%%%%%%%%%%%%%%%%%%%%%%%%%%%%%%%%%%%%%%%%%%%%%%%%%%%%%%%%%%%%%%%%%%%%%%%%%%%%%%%%%%%%%%%%%%%%%%%%%%%%%%%%%%%%%%%%%%%%%%%%%%%%%%%%%%%%%%%%%%%%%%%%%%%%%%%%%%%%%%%%%%%%%%%%%%%%%%%%%%%%%%%%%%%%%%%%%%%%%%%%%%%%%%%%%%%%%%%%%%%%%%%%%%%%%%%%%%%%%%%%%%%%%%%%%%%%%%%%%%%%%%%%%%%%%%%%%%%%%%%%%%%%%%%%%%%%%%%%%%%%%%%%%%%%%%%%%%%%%%%%%%%%%%%%%%%%%%%%%%%%%%%%%%%%%%%%%%%%%%%%%%%%%%%%%%%%%%%%%%%%%%%%%%%%%%%%%%%%%%%%%%%%%%%%%%%%%%%%%%%%%%%%%%%%%%%%%%%%%%%%%%%%%%%%%%%%%%%%%%%%%%%%%%%%%%%%%%%%%%%%%%%%%%%%%%%%%%%%%%%%%%%%%%%%%%%%%%%%%%%%%%%%%%%%%%%%%%%%%%%%%%%%%%%%%%%%%%%%%%%%%%%%%%%%%%%%%%%%%%%%%%%%%%%%%%%%%%%%%%%%%%%%%%%%%%%%%%%%%%%%%%%%%%%%%%%%%%%%%%%%%%%%%%%%%%%%%%%%%%%%%%%%%%%%%%%%%%%%%%%%%%
	
\section{Generalization to higher rank  and global charges}
\label{sec:examples}

In this section we first generalize the results obtained above to the case of a higher rank gauge group on the worldsheet, provide 
a consistent assignment of global charges and then consider a concrete example, 
in which the base manifold is a quartic  in $\mathbb{P}^{3}$.
	
\subsection{Higher rank gauge groups on the worldsheet}

As mentioned above, the whole construction of the torsion linear sigma-model can be carried on with a larger Abelian\footnote{One could 
consider general non-Abelian gauge groups, with the torsion multiplet charged under their Abelian part.} gauge group $G=U(1)^{k}$. 
For each component $U(1)_{\kappa}$, one introduces:
\begin{itemize}
\item A $(0,2)$ vector multiplet $(A_{+\kappa},A_{-\kappa})$,
\item A chiral multiplet $P_{\kappa}$,
\item A set of $r^{\kappa}+1$ Fermi multiplets $\Gamma_{a_{\kappa}}$,
\item A set of quasi-homogeneous polynomials $J_{a_{\kappa}}(\Phi_{i})$,
\end{itemize}
and a superpotential 
\begin{equation}
\mathcal{L}=\int\d\theta^{+}\, \tilde{\Gamma}_{\alpha}G^{\alpha}(\Phi_i)+\int\d\theta^{+}\ \sum_{\kappa}P_{\kappa}\ \Gamma_{a_{\kappa}}J^{a_{\kappa}}(\Phi_{i})+h.c. \, .
\end{equation}
The generalization to higher rank gauge group of the twining partition function $Z_\textsc{fy} (\tau,\bar \tau, y)$ in terms of Jeffrey-Kirwan residues is immediate, 
as explained before. Using results from reference~\cite{2015CMaPh.333.1241B} one gets the twining partition function
\begin{empheq}[box=\fbox]{multline}
Z_\textsc{fy} (\tau,\bar \tau, y)  = 
\left(-2i\pi\eta(\tau)^{2}\right)^{\text{rank}(G)} \times  \\
\sum_{u^{\star}\in\mathcal{M}_{\text{sing}}}\underset{u=u^{\star}}{\text{JK-Res}}\left(\text{Q}(u^{\star}),\eta\right)
\Bigg\{
\prod_{\Phi_{i}}\frac{i\eta(\tau)}{\vartheta_{1}(\tau|Q^\kappa _{i}u_\kappa +q_{i}^{\textsc{l}}y)}
\prod_{\Gamma_{a}}\frac{i\vartheta_{1}(\tau|Q_{a}^\kappa  u_\kappa +q^{\textsc{l}}_{a}y)}{\eta(\tau)}\ \\
\qquad \qquad \qquad \qquad \quad \sum_{\mu\in\Gamma_{\textsc{l}}^{\vee}/\Gamma_{\textsc{l}}}
\frac{\Theta_{\mu}^{\Gamma_{\textsc{l}}}(\tau| p_\textsc{m}^\kappa  u_\kappa  + \hat{p}_m y )}{\eta(\tau)^{2}}
\frac{\bar{\Theta}_{\varphi(\mu)}^{\Gamma_{R}}(\bar{\tau}|0)}{\bar{\eta}(\bar{\tau})^{2}} 
\bigwedge_{\kappa =1}^{\text{rank}(G)}\d u_{\kappa }
\Bigg\},
\label{eq:oneloopdetrank}
\end{empheq}
where one defines a vector $p_\textsc{m}^\kappa  \in \Gamma_{\textsc{l}}$ as~(\ref{eq:chargebundle}) for each $\kappa $ and 
an extra vector $\hat{p}_m$ defining the shift charge for the $U(1)_{\textsc{l}}$ global symmetry. 
To simplify the notations, $\Phi_{i}$ denotes all the chiral multiplets in the model, and $\Gamma_{a}$ all the Fermi multiplets.
Whenever gauge and global anomalies are absent no extra factors appear in the one-loop determinant, as before. 
		
One then has to choose a charge assignment for the fields which is compatible with the various anomaly cancellations, 
and gives the required value for the central charges and the rank of the spacetime gauge bundle. Let us assign the charges in the following way:
\begin{equation}
{\setlength\arraycolsep{12pt}
\begin{array}{|c||c|c|c|c|c|}
\hline
&\Phi_{i}&P_{\kappa}&\tilde{\Gamma}_{\alpha}&\Gamma_{a_{\kappa}}&\Theta\\
\hline\hline
U(1)_{\iota}&Q_{i}^{\iota}&-d^{\iota}_{\kappa}&Q_{\alpha}^{\iota}&Q_{a_{\kappa}}^{\iota}&M_{\ell}^{\iota}\\
\hline	
U(1)_{\textsc{l}}&0&1&0&-1&0\\
\hline		
\end{array}
\label{setofcharges}
}
\end{equation}
One sees that whenever the tadpole conditions	
\begin{equation}
\label{eq:tadrank}	
\sum_{i}Q_{i}^{\epsilon}Q_{i}^{\epsilon'}+\sum_{\kappa}d_{\kappa}^{\epsilon}
d_{\kappa}^{\epsilon'}-\sum_{\alpha}Q_{\alpha}^{\epsilon}Q_{\alpha}^{\epsilon'}
-\sum_{\kappa,a_{\kappa}}Q_{a_{\kappa}}^{\epsilon}Q_{a_{\kappa}}^{\epsilon'}
-\langle p_{\textsc{m}}^{\epsilon},p_{\textsc{m}}^{\epsilon'}\rangle=0\ \ \ \ 
\forall \epsilon,\epsilon'\in\{1,2,...,k\}\, ,
\end{equation}
are satisfied, then the various local and global anomalies vanish.  Furthermore, the rank $r$ of the holomorphic vector bundle is then given by 
construction by $r=\sum_{\kappa}r^{\kappa}$, the central charges take the appropriate values $(c,\bar c)= (6+r,9)$ and the left and right global $U(1)$ current decouple, 
see~\cite{2011JHEP...03..045A} for details.\footnote{Compared to the work of Adams and Lapan, we have shifted all the $U(1)_L$ charges using the gauge shift 
$\tilde{u}_{\kappa}:=u_{\kappa}-\sum_{\iota}(d^{-1})^{\iota}_{\kappa}y$. 
Our choice of charges turns out to be more appropriate in order to discuss the link with the geometrical formula of \
section~\ref{sec:geom}.} Finally this choice of charge is consistent with a space-time gauge bundle having vanishing first Chern class.

This choice of global charges implies that in the geometrical "phase" of the torsion GLSM, which corresponds to taking the residues 
at the poles of the chiral multiplets $\Phi_{i}$ (i.e. points $u^{\star}$ such that $Q^{\kappa}_{i}u_{\kappa}^{\star} \in \mathbb{Z}+\tau\mathbb{Z}$), 
the contribution from the torsion multiplet has no $y$-dependence, in keep with the geometrical formula that we define in section~\ref{sec:geom}.
 From a geometrical point of view, the meaning of this absence of $y$-dependence is that the torus fiber should not contribute to the rank of the holomorphic vector bundle. This assertion becomes transparent when we examine the modular behavior of $Z_{\textsc{fy}}$.

\subsubsection*{Modular transformations}
Let us denote by $d$ the complex dimension of the base, $k$ the rank of the worldsheet gauge group and $r$ the rank of the space-time holomorphic vector bundle. 
Using the results of appendix~\ref{app:theta}, the behavior of $\Sigma_{\text{1-loop}}$ under the $SL(2,\mathbb{Z})$ modular transformations is straightforward.
Under a modular T-transformation $\tau \mapsto \tau+1$, the torsion multiplet contribution is by itself invariant.\footnote{
The isometry $\phi$ preserving the bilinear form, $\langle \mu,\mu\rangle_{\Gamma_L} = \langle \phi(\mu),\phi(\mu)\rangle_{\Gamma_R}$.} The remaining 
contribution comes from the base, and gives:
\begin{equation}
\Sigma_{\text{1-loop}}(\tau+1,\bar{\tau}+1,y,u_{\kappa})=e^{-\frac{i\pi}{6}(d-r)}\Sigma_{\text{1-loop}}(\tau,\bar{\tau},y,u_{\kappa})\, .
\end{equation}
Under an S-transformation $\tau \mapsto -1/\tau$, one finds the following transformation rule:
\begin{align}
\label{eq:modtransf}
&\Sigma_{\text{1-loop}}\left(-\frac{1}{\tau},-\frac{1}{\bar{\tau}},\frac{y}{\tau},\frac{u_{\kappa}}{\tau}\right)=\nonumber\\
&\ \ \ \ i^{d-r}\exp\left\{-\frac{i\pi}{\tau}\left(\sum_{\Phi_{i}}v_{i}^{\ 2}+
\sum_{P_{\kappa}}v_{\kappa}^{\ 2}-\sum_{\tilde{\Gamma}_{\alpha}}v_{\alpha}^{\ 2}-\sum_{\Gamma_{a_{\kappa}}}v_{a_{\kappa}}^{\ 2}-\langle v,v\rangle_{\Gamma_L}\right)\right\}
\Sigma_{\text{1-loop}}(\tau,\bar{\tau},y,u_{\kappa})\, ,
\end{align}
where
\begin{subequations}
\begin{align}
v_{i}&=Q_{i}^{\epsilon}u_{\epsilon}\, ,\\
v_{\kappa}&=-d_{\kappa}^{\epsilon}u_{\epsilon}+y\, ,\\
v_{\alpha}&=Q_{\alpha}^{\epsilon}u_{\epsilon}\, ,\\
v_{a_{\kappa}}&=Q_{a_{\kappa}}^{\epsilon}u_{\epsilon}-y\, ,\\
v&=p_{\textsc{m}}^{\epsilon}u_{\epsilon}\, .
\end{align}
\end{subequations}
The charge assignement given by~(\ref{setofcharges}) was precisely designed such that the gauge and global anomalies vanish provided that 
the tadpole conditions~(\ref{eq:tadrank}) hold. One gets then:
\begin{equation}
\Sigma_{\text{1-loop}}\left(-\frac{1}{\tau},-\frac{1}{\bar{\tau}},\frac{y}{\tau},\frac{u_{\kappa}}{\tau}\right)=i^{d-r}\exp\left[\frac{2i\pi}{\tau}\frac{r}{2}y^{2}\right]\Sigma_{\text{1-loop}}(\tau,\bar{\tau},y,u_{\kappa})\, .
\end{equation}
One concludes that, though non-holomorphic in $\tau$, the twining partition function $Z_\textsc{fy}$ transforms as a weak Jacobi form of index $\frac{r}{2}$ and weight zero. 
This result will be derived again starting from the geometrical formula that we provide in section \ref{sec:geom}.

\subsection{Example of the quartic}

We illustrate here the formula giving the twining 
partition function $Z_{\textsc{fy}}$ of Fu-Yau compactifications in terms of the torsion GLSM data with a simple example, 
namely a quartic hypersurface in $\mathbb{P}^{3}$ with a rank four gauge bundle~\cite{2011JHEP...03..045A}. Following (\ref{setofcharges}), let the charges for the base be: 
\begin{equation}
{\setlength\arraycolsep{8pt}
\begin{array}{|c||c|c|c|c||c|}
\hline  
&\Phi^{i=1,...,4}&P&\tilde{\Gamma}&\Gamma_{a=1,...,5}&\Theta\\
\hline\hline 
U(1)&1&-5&-4&1&M_{\ell}\\
\hline 
U(1)_{\textsc{l}}&0&1&0&-1&0\\
\hline
\end{array}}\, ,
\end{equation}
with, in addition, the moduli $(T,U)$ and the topological charge $M$ of the torus fiber chosen such that the tadpole condition~(\ref{eq:tadpole}) is satisfied. 
The full one-loop determinant writes
\begin{align}
\Sigma_{\text{1-loop}}=&\left[-2i\pi\eta(\tau)^{2}\right]\left[i\frac{\eta(\tau)}{\vartheta_{1}\left(\tau\left|u\right.\right)}\right]^{4}
\left[i\frac{\eta(\tau)}{\vartheta_{1}\left(\tau\left|-5u+y\right.\right)}\right]
\left[i\frac{\vartheta_{1}\left(\tau\left|-4u\right.\right)}{\eta(\tau)}\right]
\left[i\frac{\vartheta_{1}\left(\tau\left|u-y\right.\right)}{\eta(\tau)}\right]^{5}\times\nonumber\\
&\times\left[\sum_{\mu\in\Gamma_{\textsc{l}}^{\vee}/\Gamma_{\textsc{l}}}
\frac{\Theta_{\mu}^{\Gamma_{\textsc{l}}}\left(\tau\left|p_\textsc{m}u\right.\right)}{\eta(\tau)^{2}}
\frac{\bar{\Theta}_{\varphi(\mu)}^{\Gamma_{R}}(\bar{\tau}|0)}{\bar{\eta}(\bar{\tau})^{2}}\right]\d u\, .
\end{align}

\subsubsection*{Landau-Ginzburg phase}
One  can first provide the result in a form that one would obtain by a direct computation in the Landau-Ginzburg regime 
of the base GLSM. For this purpose one selects the set of poles $\mathcal{M}_{\text{sing}}^{-}=
\left\{\left.u=-\frac{k+\tau l-y}{5}\ \right|k,l\in\llbracket0,4\rrbracket\right\}$.
Plugging the one-loop determinant into the contour integral $\frac{1}{2i\pi}\oint$ leads to:
\begin{align}
Z_{\textsc{fy}}(\tau,\bar{\tau},y)=-\frac{i}{\eta(\tau)\bar{\eta}(\bar{\tau})^{2}}
\sum_{k,l=0}^{4}\oint_{u=-\frac{k+\tau l-y}{5}}\d u\ &\frac{\vartheta_{1}
\left(\tau\left|u-y\right.\right)^{5}}{\vartheta_{1}\left(\tau\left|u\right.\right)^{4}}
\frac{\vartheta_{1}\left(\tau\left|-4u\right.\right)}{\vartheta_{1}\left(\tau\left|-5u+y\right.\right)}\times\nonumber\\
&\times\sum_{\mu\in\Gamma_{\textsc{l}}^{\vee}/\Gamma_{\textsc{l}}}\Theta_{\mu}^{\Gamma_{\textsc{l}}}
\left(\tau\left|p_\textsc{m}u\right.\right)\bar{\Theta}_{\varphi(\mu)}^{\Gamma_{R}}(\bar{\tau}|0)\, .
\end{align}
Evaluating the residues, one has
\begin{align}
Z_{\textsc{fy}}(\tau,\bar{\tau},y)=\frac{1}{5\eta(\tau)^{4}\bar{\eta}(\bar{\tau})^{2}}
\sum_{k,l=0}^{4}&(-1)^{k+l}e^{i\pi l^{2}\tau}\frac{\vartheta_{1}\left(\tau\left|-\frac{k+\tau l}{5}
-\frac{4y}{5}\right.\right)^{5}}{\vartheta_{1}\left(\tau\left|-\frac{k+\tau l}{5}
+\frac{y}{5}\right.\right)^{4}}\vartheta_{1}\left(\tau\left|\frac{4(k+\tau l)}{5}-\frac{4y}{5}\right.\right)\times\nonumber\\
&\ \ \ \ \ \ \times\sum_{\mu\in\Gamma_{\textsc{l}}^{\vee}/\Gamma_{\textsc{l}}}\Theta_{\mu}^{\Gamma_{\textsc{l}}}
\left(\tau\left|\left(-\frac{k+\tau l}{5}+\frac{y}{5}\right)p_{\textsc{m}}\right.\right)\bar{\Theta}_{\varphi(\mu)}^{\Gamma_{R}}(\bar{\tau}|0)\, .
\end{align}

\subsubsection*{Geometrical phase}

An expression corresponding to a direct computation in the geometrical formulation of the index, see section~\ref{sec:geom}, is obtained by 
considering the contribution of the pole $u=0$, which is of order 4.\footnote{As was noted earlier, the expressions that one gets by 
choosing the poles in $\mathcal{M}_{\text{sing}}^{-}$ (Landau-Ginzburg picture) or in $\mathcal{M}_{\text{sing}}^{+}$ (geometrical picture) 
coincide.} Plugging the one loop determinant into the contour 
integral $-\frac{1}{2i\pi}\oint$ leads to the expression
\begin{align}
\label{eq:geomquartic}
Z_{\textsc{fy}}(\tau,\bar{\tau},y)=\frac{i}{\eta(\tau)\bar{\eta}(\bar{\tau})^{2}}
\oint_{u=0}\d u\ &\frac{\vartheta_{1}\left(\tau\left|u-y\right.\right)^{5}}{\vartheta_{1}
\left(\tau\left|u\right.\right)^{4}}\frac{\vartheta_{1}
\left(\tau\left|-4u\right.\right)}{\vartheta_{1}\left(\tau\left|-5u+y\right.\right)}\times\nonumber\\
&\ \ \ \ \ \ \times \sum_{\mu\in\Gamma_{\textsc{l}}^{\vee}/\Gamma_{\textsc{l}}}
\Theta_{\mu}^{\Gamma_{\textsc{l}}}\left(\tau\left|p_\textsc{m}u\right.\right)\bar{\Theta}_{\varphi(\mu)}^{\Gamma_{R}}(\bar{\tau}|0)\, .
\end{align}

To conclude this section, let us consider one specific consistent choice of two-torus fiber. To illustrate what happens for a non-orthogonal torus with non-vanishing $B$-field, one takes 
the Wess-Zumino-Witten theory $\widehat{\mathfrak{su}(3)}_{1}$.\footnote{This is a special case of the construction discussed in~\cite{Adams:2009av}.} 
It corresponds to a $c=2$ toroidal rational CFT with $T$ and $U$ both equal to the cubic root of unity $j=\exp\left(\frac{2i\pi}{3}\right)$, satisfying the quadratic equation 
$j^{2}+j+1=0$. Hence $T$ and $U$ belong to the same imaginary quadratic number field $\mathbb{Q}(\sqrt{-3})$. 

A consistent choice of topological charge is given by $M_{1}=M_{2}=2$, corresponding to the following vector in the root lattice $\mathfrak{su}(3)\simeq A_{2}$:
\begin{equation}
p_{\textsc{m}}=\sqrt{2}\begin{pmatrix} 1\\ \sqrt{3}\end{pmatrix}\label{su(3)_charge}\, ,
\end{equation}
written in an orthonormal basis. The root lattice $A_{2}$ has discriminant group $A^{\vee}_{2}/A_{2}\simeq\mathbb{Z}_{3}$. Hence, in terms of the $SU(3)$ theta functions
\begin{equation}
\Theta_{\mu}^{A_{2}}(\tau|\lambda)=\sum_{\gamma\in A_{2}+\mu}q^{\frac{1}{2}\langle\gamma,\gamma\rangle}e^{2i\pi\langle\gamma,\lambda\rangle}\, ,
\end{equation}
one has the following twining partition function
\begin{align}
Z_{\textsc{fy}}(\tau,\bar{\tau},y)=\frac{1}{5\eta(\tau)^{4}\bar{\eta}(\bar{\tau})^{2}}\sum_{k,l=0}^{4}&(-1)^{k+l}
e^{i\pi l^{2}\tau}\frac{\vartheta_{1}\left(\tau\left|-\frac{k+\tau l}{5}-\frac{4y}{5}\right.\right)^{5}}{\vartheta_{1}
\left(\tau\left|-\frac{k+\tau l}{5}+\frac{y}{5}\right.\right)^{4}}\vartheta_{1}
\left(\tau\left|\frac{4(k+\tau l)}{5}-\frac{4y}{5}\right.\right)\times\nonumber\\
&\times\sum_{\mu\in\mathbb{Z}_{3}}\Theta_{\mu}^{A_{2}}
\left(\tau\left|\left(-\frac{k+\tau l}{5}+\frac{y}{5}\right)p_{\textsc{m}}\right.\right)\bar{\Theta}_{\varphi(\mu)}^{A_{2}}(\bar{\tau}|0)\, , 
\end{align}
that we have evaluated in the Landau-Ginzburg phase.
		
Notice that this model is non-supersymmetric in spacetime, as the primitivity condition~\eqref{eq:primitivity} is not satisfied, 
the two-form $\omega$ being necessarily proportional to the K\"ahler form of the base $J_{\text{K3}}$ (cf. \cref{appendixFuYau}). Supersymmetric 
examples are easily obtained with higher rank worldsheet gauge groups; instead of dealing with such examples in detail, we will provide 
below a formulation of the index which is independent of the choice of GLSM. 		
		
%%%%%%%%%%%%%%%%%%%%%%%%%%%%%%%%%%%%%%%%%%%%%%%%%%%%%%%%%%%%%%%%%%%%%%%%%%%%%%%%%%%%%%%%%%%%%%%%%%%%%%%%%%%%%%%%%%%%%%%%%%%%%
%%%%%%%%%%%%%%%%%%%%%%%%%%%%%%%%%%%%%%%%%%%%%%%%%%%%%%%%%%%%%%%%%%%%%%%%%%%%%%%%%%%%%%%%%%%%%%%%%%%%%%%%%%%%%%%%%%%%%%%%%%%%%
%%%%%%%%%%%%%%%%%%%%%%%%%%%%%%%%%%%%%%%%%%%%%%%%%%%%%%%%%%%%%%%%%%%%%%%%%%%%%%%%%%%%%%%%%%%%%%%%%%%%%%%%%%%%%%%%%%%%%%%%%%%%%
%%%%%%%%%%%%%%%%%%%%%%%%%%%%%%%%%%%%%%%%%%%%%%%%%%%%%%%%%%%%%%%%%%%%%%%%%%%%%%%%%%%%%%%%%%%%%%%%%%%%%%%%%%%%%%%%%%%%%%%%%%%%%
		
\section{A geometrical formula for the non-holomorphic genus}
\label{sec:geom}

The elliptic genus of a complex manifold $M$ of dimension $d$, of holomorphic tangent bundle $\mathcal T_M$, with a holomorphic vector 
bundle $\mathcal V$ of rank $r$ over it, can be defined independently of its 
realization as the target space of a $(0,2)$ superconformal field theory. One defines the formal power series 
\begin{equation}\label{eq:formalbundle}
\mathbb{V}_{q,w} = \bigotimes_{n=0}^\infty \bigwedge\nolimits_{-w q^n} \mathcal V^\star \otimes  
\bigotimes_{n=1}^\infty \bigwedge\nolimits_{-w^{-1} q^n} \mathcal  V \otimes 
\bigotimes_{n=1}^\infty S_{q^n} \mathcal T^\star_M \otimes 
\bigotimes_{n=1}^\infty S_{q^n} \mathcal T_M\, ,
\end{equation}
where 
\begin{equation}
\bigwedge\nolimits_{t} \mathcal V = 1+ t\, \mathcal V + t^2\, \bigwedge\nolimits^2 \mathcal V + \cdots \ , \quad 
S_t \mathcal T_M = 1+ t\,  \mathcal T_M + t^2 \, S^2 \, \mathcal T_M + \cdots\, ,
\end{equation}
$\bigwedge\nolimits^k$ and $S^k$ being respectively  the $k$-th exterior product and the $k$-th symmetric product. 
The elliptic genus corresponding to this bundle is defined as follows:
\begin{equation}
Z_\textsc{ell} (M,\mathcal V|\tau,y) = q^{\frac{r-d}{12}} w^{-\frac{r}{2}}\int_M 
\text{ch}\, (\mathbb{V}_{q,w})\, \text{td} (\mathcal T_M)\, ,
\end{equation}
where $\text{ch}\, (\mathbb{V}_{q,w})$ is the Chern character of the formal power series $\mathbb{V}_{q,w}$ and 
$\text{td}\, (\mathcal T_M)$ the Todd class of the tangent bundle. 
Considering that $M$ is a Calabi-Yau manifold,  that $\mathcal V$ has vanishing first Chern class, and that the 
anomaly condition $\text{ch}_2 (\mathcal{T}_M) = \text{ch}_2 ( \mathcal V)$ is satisfied (that is, we consider an 
anomaly-free heterotic Calabi-Yau compactification), the elliptic genus is a weak Jacobi form of weight 0 and index $r/2$.

This geometrical formula has been checked against $(0,2)$ Landau-Ginzburg results in \cite{Kawai:1993jk}, and directly 
compared with the results of supersymmetric localization for $(2,2)$ GLSMs in~\cite{2014LMaPh.104..465B,2015CMaPh.333.1241B}, 
building on previous works in the physical and mathematical 
literature~\cite{Kawai:1994np,zbMATH02150990,doi:10.1142/S0129167X05003259,doi:10.1142/S0129167X11007008,zbMATH05291408}.

In the present 
context, there is a natural generalization of this geometrical formulation of the Calabi-Yau elliptic genera, 
defining a non-holomorphic genus for a two-torus bundle over a K3 surface 
$\mathcal{S}$, $T^2 \hookrightarrow \mathcal{M} \stackrel{\pi}{\to} \mathcal{S}$, 
endowed with a rank $r$ gauge bundle $\mathcal V$. The relevant geometrical data of such 
non-K\"ahler manifold is given by 
\begin{itemize}
\item The holomorphic tangent bundle $\mathcal{T}_\mathcal{S}$ over the base, with $c_1 (\mathcal{T}_\mathcal{S}) = 0$, 
\item A rank $r$ holomorphic vector bundle $\mathcal V$ over $\mathcal{S}$, with $c_1 (\mathcal{V}) = 0$, whose pullback provides the 
gauge bundle of the compactification on $\mathcal{M}$,\footnote{
Considering that the holomorphic gauge bundle has vanishing first Chern class is not mandatory for 
getting consistent heterotic compactifications; it is enough that 
$c_1 (\mathcal V)\in H^{2}(\mathcal{S},2\mathbb{Z})$ 
($i.e.$ vanishing of the second Stieffel-Whitney class) to ensure that the bundle admits spinors.} 
\item A rational Narain lattice $\Gamma (T,U)$ with $T, U \in \mathbb{Q}(\sqrt{D})$, or equivalently a 
triple\newline $[\Gamma_{\textsc{l}},\Gamma_{\textsc{r}}, \phi]$ defining a $c=2$ toroidal rational CFT,
\item A pair of anti-self-dual two-forms $\omega_1$ and $\omega_2$ in $H^{2}(\mathcal{S},\mathbb{Z})\cap 
\Lambda^{1,1} \mathcal{T}_\mathcal{S}^\star$.
\end{itemize}

We define then the non-holomorphic genus of the Fu-Yau compactification in terms of this data 
as:\footnote{Notice that in eq.~(\ref{eq:nonholgen}) the tangent bundle of the base $\mathcal{S}$, rather than of the total space 
$\mathcal{M}$, appears. This makes sense as the Chern classes of $\mathcal{T}_\mathcal{M}$ are 'horizontal', $i.e.$ with no 
components along the torus fiber.}
\begin{equation}
\label{eq:nonholgen}
\boxed{
Z_\textsc{fy} (\mathcal{M},\mathcal V,\omega|\tau,\bar \tau,y) = 
q^{\frac{r-2}{12}} w^{-\frac{r}{2}}\int_\mathcal{S} \text{ch}\, (\mathbb{V}_{q,w})\, \text{td}\, (\mathcal{T}_\mathcal{S}) 
\sum_{\mu\in\Gamma_{\textsc{l}}^{\vee}/\Gamma_{\textsc{l}}}
\frac{\Theta_{\mu}^{\Gamma_{\textsc{l}}}\left(\tau\left|\frac{p_{\omega}}{2i\pi}\right.\right)}{\eta(\tau)^{2}}
\frac{\bar{\Theta}_{\varphi(\mu)}^{\Gamma_{\textsc{r}}}(\bar{\tau}|0)}{\bar{\eta}(\bar{\tau})^{2}}
}\, ,
\end{equation}
where the two-component vector $p_\omega$ valued in $H^2 (\mathcal{S}) \times H^2 (\mathcal{S})$ reads, taking  
an orthonormal basis on $\Gamma_{\textsc{l}}$:
\begin{equation}
p_{\omega}=\sqrt{\frac{2U_{2}}{T_{2}}}\begin{pmatrix}\omega_{1}+T_{1}\, 
\omega_{2}\\T_{2}\, \omega_{2}\end{pmatrix}\, ,\label{vectorThetaforms}
\end{equation} 
which generalizes eq.~(\ref{vectorTheta}). This vector belongs to a formal extension of 
the left momentum lattice $\Gamma_{\textsc{l}}$, which is now a module over $H^2 (\mathcal{S},\mathbb{Z})$.

A more explicit expression can be obtained using the splitting principle. Let $c(\mathcal{T}_\mathcal{S})=\prod_{i=1}^{2}(1+\nu_{i})$ 
and $c(\mathcal V)=\prod_{a=1}^{r}(1+\xi_{a})$ denote the total Chern classes of the respective bundles. We have then 
\begin{equation}
Z_\textsc{fy} (\mathcal{M},\mathcal V,\omega|\tau,\bar \tau,y) =\int_{\mathcal{S}}G(\tau,\bar{\tau},y,\nu,\xi,p_\omega)\, 
,\label{eq:geometric_formula}
\end{equation}
where
\begin{equation}
\label{eq:integrand}
G(\tau,\bar{\tau},y,\nu,\xi,p_\omega)=\prod_{a=1}^{r}\frac{i\vartheta_{1}(\tau\left|\frac{\xi_{a}}{2i\pi}-y\right.)}{\eta(\tau)}
\prod_{i=1}^{2}\frac{\eta(\tau)\nu_{i}}{i\vartheta_{1}(\tau\left|\frac{\nu_{i}}{2i\pi}\right.)}
\sum_{\mu\in\Gamma_{\textsc{l}}^{\vee}/\Gamma_{\textsc{l}}}
\frac{\Theta_{\mu}^{\Gamma_{\textsc{l}}}\left(\tau\left|\frac{p_{\omega}}{2i\pi}\right.\right)}{\eta(\tau)^{2}}
\frac{\bar{\Theta}_{\varphi(\mu)}^{\Gamma_{\textsc{r}}}(\bar{\tau}|0)}{\bar{\eta}(\bar{\tau})^{2}}\, .
\end{equation}
		
\subsection{Modular properties}
\label{sec:modularprop}
The behaviour of $G(\tau,\bar{\tau},y,\nu,\xi,p_\omega)$ under $PSL(2,\mathbb{Z})_{\tau}$ is easily derived using 
the results of appendix \ref{app:theta}. Under the T-transformation $\tau\mapsto \tau+1$, one gets 
\begin{equation}
G(\tau+1,\bar{\tau}+1,y,\nu,\xi,p_\omega)=e^{-\frac{i\pi}{6}(2-r)}G(\tau,\bar{\tau},y,\nu,\xi,p_\omega)\, ,
\end{equation}
exactly as in the GLSM computation of section~\ref{sec:examples}.

The contributions of the holomorphic vector bundle $\mathcal V$ and of the tangent bundle  $T_\mathcal{S}$ 
to $G(\tau,\bar{\tau},y,\nu,\xi,p_\omega)$ behave under an $S$-transformation 
$\tau \mapsto -1/\tau$ as
\begin{subequations}
\begin{align}
\label{eq:vectortrans}
\prod_{a=1}^{r}\frac{\vartheta_{1}\left(-\frac{1}{\tau}\left|
\frac{\xi_{a}/2i\pi-y}{\tau}\right.\right)}{\eta\left(-\frac{1}{\tau}\right)}&=
\prod_{a=1}^{r}\left\{-ie^{\frac{i\pi}{\tau}\left(\frac{\xi_{a}}{2i\pi}-y\right)^{2}}
\frac{\vartheta_{1}(\tau\left|\frac{\xi_{a}}{2i\pi}-y\right.)}{\eta(\tau)}\right\}\, , \\
\prod_{i=1}^{2}\frac{\eta\left(-\frac{1}{\tau}\right)\frac{\nu_{i}}{\tau}}{\vartheta_{1}
\left(-\frac{1}{\tau}\left|\frac{\nu_{i}/2i\pi}{\tau}\right.\right)} &=
\frac{1}{\tau^{2}}\prod_{i=1}^{2}\left\{ie^{-\frac{i\pi}{\tau} \left(\frac{\nu_{i}}{2i\pi}\right)^{\ 2}}
\frac{\eta(\tau)\nu_{i}}{\vartheta_{1}(\tau\left|\frac{\nu_{i}}{2i\pi}\right.)}\right\}\, .
\label{eq:tangenttrans}
\end{align}
\end{subequations}
One recognizes on the right-hand side of~(\ref{eq:vectortrans},\ref{eq:tangenttrans}) the second Chern characters of 
the vector bundle and of the tangent bundle:
\begin{subequations}
\begin{align}
\text{ch}_{2}(\mathcal V)&=\frac{1}{2}\, \text{tr}\, \left(\frac{i}{2\pi}F\right)^{2}=\frac{1}{2}\sum_{a=1}^{r}\xi_{a}^{\ 2}\, , \\
\text{ch}_{2}(\mathcal{T}_\mathcal{S})&=\frac{1}{2}\, \text{tr}\, \left(\frac{i}{2\pi}R\right)^{2}=\frac{1}{2}\sum_{i=1}^{2}\nu_{i}^{\ 2}\, ,
\end{align}
\label{eq:chernchar}
\end{subequations}
Combining these expressions with the contribution from the torus fiber, obtained using the modular transformation of theta-functions 
given by~(\ref{eq:thetatrans}) in appendix~\ref{app:theta}, one gets
\begin{multline}
\label{modular_behaviour}
G\left(-\frac{1}{\tau},-\frac{1}{\bar{\tau}},\frac{y}{\tau},\frac{\nu}{\tau},\frac{\xi}{\tau},\frac{p_\omega}{\tau}\right)= \\ 
-(-i)^{r}\tau^{-2}e^{2i\pi\frac{r}{2}
\frac{y^{2}}{\tau}+\frac{2i\pi}{\tau} \frac{\text{ch}_{2}(\mathcal V)-\text{ch}_{2}(\mathcal{T}_\mathcal{S})}{(2i\pi)^2}}
e^{\frac{i\pi}{\tau} \frac{\langle p_\omega ,p_\omega \rangle }{(2i\pi)^2}}
G(\tau,\bar{\tau},y,\nu,\xi,p_\omega)\, ,
\end{multline}
with 
\begin{equation}
\langle p_\omega ,p_\omega \rangle  = \frac{2U_2}{T_2} (\omega_1 + T\omega_2)\wedge (\omega_1 + \bar{T}\omega_2)= 
-\frac{2U_2}{T_2} \omega \wedge \star_{\scriptscriptstyle \mathcal{S}}\,  \bar{\omega}\, ,
\end{equation}
using the anti-self-duality property of the complex two-form $\omega$. 

In conclusion we obtain that $G$, 
although non-holomorphic in $\tau$, transforms as a Jacobi form of weight $-2$ and index $\frac{r}{2}$, whenever 
the anomaly cancellation condition 
\begin{equation}
\text{ch}_{2}(\mathcal V)-\frac{U_{2}}{T_{2}}\omega\wedge\star_{\scriptscriptstyle \mathcal{S}}\, \bar{\omega}=\text{ch}_{2}(\mathcal{T}_\mathcal{S})
\label{eq:geom_anomaly}
\end{equation}
is satisfied. This condition corresponds exactly to the Bianchi identity for Fu-Yau compactifications, see 
appendix \ref{appendixFuYau}. After integrating $G$ over $\mathcal{S}$, through a Taylor expansion to 
second order in the differential forms, the non-holomorphic genus $Z_\textsc{fy}$ transforms then as a Jacobi form of weight zero 
and index $\frac{r}{2}$. 

Let us remark here that the expression of the geometrical formula~(\ref{eq:nonholgen}) is pretty much fixed by its modular 
behavior. In particular, the absence of $y$-dependence in the torus fibration contribution, 
hence the absence of shift in the rank of bundle from the latter under modular transformations, 
is compatible with the result that one obtains when evaluating the twining 
partition function  of the gauged linear sigma-model description in the geometrical phase.

We indeed expect that the geometrical formula~(\ref{eq:nonholgen}) and the GLSM formula~(\ref{eq:oneloopdetrank}) for the 
non-holomorphic genus $Z_\textsc{fy}(\tau,\bar\tau, y)$ coincide. A general mathematical proof should follow from a 	
natural generalization of the arguments in~\cite{doi:10.1142/S0129167X05003259,doi:10.1142/S0129167X11007008}, 
first to $(0,2)$ Calabi-Yau examples and second to the Fu-Yau geometries under consideration in the present article. We 
provide below a proof of this statement in a simple case.

\subsection{Proof of the geometrical formula for the quartic}

We consider here the same example as in section~\ref{sec:examples}, namely a Fu-Yau manifold $\mathcal{M}$ 
based on a quartic hypersurface $\mathcal{S}$ in $\mathbb{P}^3$, with a monad rank four holomorphic 
vector bundle $\mathcal{V}$. Let us first define: 
\begin{multline}
\text{ch}\, (\mathbb{V}_{q,w}) \, \text{Td}\, (\mathcal{S})  =\\ 
\underbrace{\text{ch}\, \left(\bigotimes_{n=0}^\infty \bigwedge\nolimits_{-w q^n} \mathcal V^\star \otimes  
\bigotimes_{n=1}^\infty \bigwedge\nolimits_{-w^{-1} q^n} \mathcal  V\right)}_{f(\mathcal{V})} \, 
\underbrace{\text{ch}\, \left( \bigotimes_{n=1}^\infty S_{q^n} \mathcal{T}^\star_{\mathcal{S}} \otimes 
\bigotimes_{n=1}^\infty S_{q^n} \mathcal{T}_\mathcal{S} \right) \, \text{Td}\, (\mathcal{S})}_{g (\mathcal{T}_\mathcal{S})}\, .
\end{multline}
Using the adjunction formula for the hypersurface $\mathcal{S}$ in $\mathbb{P}^3$, the properties 
of the Chern character and of the Todd class one finds the relation:
\begin{equation}
g \left( \mathcal{T}_{\mathbb{P}^3}\big|_\mathcal{S} \right) = 
g \left( \mathcal{T}_\mathcal{S} \right) g \left( \mathcal{O}_{\mathbb{P}^3} (4) \big|_{\mathcal{S}} \right) \, .
\end{equation}
Since $c (\mathcal{T}_{\mathbb{P}^3}) = (1+H)^4$ and $c( \mathcal{O}_{\mathbb{P}^3} (4)) = 1+4H$, $H$ being the hyperplane class, 
one gets: 
\begin{equation}
\frac{4H \eta (\tau)}{ i\vartheta_1 (\tau|\frac{4H}{2i\pi})} \Bigg|_{\mathcal{S}}\, 
q^{-\frac{1}{6}}\, g \left( \mathcal{T}_\mathcal{S} \right) = 
\left(\frac{H \eta (\tau)}{ i\vartheta_1 (\tau|\frac{H}{2i\pi})}\right)^4 \Bigg|_{\mathcal{S}}\eta(\tau)^2\, . 
\end{equation}
We evaluate now the contribution $f(\mathcal{V})$ coming from the holomorphic vector bundle. 
This rank four bundle is defined by the short exact sequence: 
\begin{equation}
0 \longrightarrow \mathcal{V} \stackrel{\iota}{\longrightarrow} 5 \mathcal{O}_{\mathbb{P}^3} (1)  \big|_{\mathcal{S}}
\stackrel{\otimes J^a}{\longrightarrow} 
\mathcal{O}_{\mathbb{P}^3} (5) \big|_{\mathcal{S}} \longrightarrow 0\, ,
\end{equation}
and can be viewed as the restriction to $\mathcal{S}$ of a vector bundle 
over $\mathbb{P}^3$ defined by a similar sequence. The Chern roots 
$\{ \xi_a, a=1,\ldots,4 \}$ of the latter satisfy then the relation 
\begin{equation}
(1+5H) \prod_{a=1}^4 (1+\xi_a) = (1+H)^5 \, ,
\end{equation}
from which we deduce that 
\begin{equation}
\frac{i\vartheta_1 (\tau| \frac{5H}{2i\pi}-y)}{\eta(\tau)}\, q^{\frac{1}{3}} w^{-2} \, 
f (\mathcal{V})   = \left(\frac{i\vartheta_1 (\tau| \frac{H}{2i\pi}-y)}{\eta(\tau)}\right)^5  \, .
\end{equation}
Regarding the torus bundle, the vector $p_\omega$ is proportional to the hyperplane class in this example: 
\begin{equation}
p_\omega = p_\textsc{m} H = \sqrt{\frac{2U_{2}}{T_{2}}}\begin{pmatrix}M_{1}+T_{1}M_{2}\\T_{2}M_{2}\end{pmatrix} H \, .
\end{equation} 
Putting all pieces together we obtain for the non-holomorphic genus: 
\begin{multline}
Z_\textsc{fy} (\mathcal{M},\mathcal V,\omega|\tau,\bar \tau,y) = 
q^{\frac{1}{6}} w^{-2}\int_\mathcal{S} f (\mathcal{V}) \, g (\mathcal{T}_\mathcal{S})
\sum_{\mu\in\Gamma_{\textsc{l}}^{\vee}/\Gamma_{\textsc{l}}}
\frac{\Theta_{\mu}^{\Gamma_{\textsc{l}}}\left(\tau\left|p_\textsc{m} \frac{H}{2i\pi}\right.\right)}{\eta(\tau)^{2}}
\frac{\bar{\Theta}_{\varphi(\mu)}^{\Gamma_{\textsc{r}}}(\bar{\tau}|0)}{\bar{\eta}(\bar{\tau})^{2}} \\ 
= \int_{\mathbb{P}^3} \text{c}_1\, \big( \mathcal{O}_{\mathbb{P}^3 } (4) \big)\,
\left(\frac{i\vartheta_1 (\tau| \frac{H}{2i\pi}-y)}{\eta(\tau)}\right)^5
\frac{\eta(\tau)}{i\vartheta_1 (\tau| \frac{5H}{2i\pi}-y)}
\left(\frac{H \eta (\tau)}{ i\vartheta_1 (\tau|\frac{H}{2i\pi})}\right)^4
\frac{i\vartheta_1 (\tau|\frac{4H}{2i\pi})}{4H \eta (\tau)}\ \eta(\tau)^2\ \times \\
\sum_{\mu\in\Gamma_{\textsc{l}}^{\vee}/\Gamma_{\textsc{l}}}
\frac{\Theta_{\mu}^{\Gamma_{\textsc{l}}}\left(\tau\left|p_\textsc{m} \frac{H}{2i\pi}\right.\right)}{\eta(\tau)^{2}}
\frac{\bar{\Theta}_{\varphi(\mu)}^{\Gamma_{\textsc{r}}}(\bar{\tau}|0)}{\bar{\eta}(\bar{\tau})^{2}}\, .
\end{multline}
Finally, using a residue theorem on $\mathbb{P}^3$:
\begin{equation}
\int_{\mathbb{P}^3} H^4  \phi \left( H\right) = \oint_{u=0} {\rm d}u\, \phi (2i\pi u)\, ,
\end{equation}
we get exactly the same formula as  eq.~(\ref{eq:geomquartic}), the outcome of the torsion GLSM computation.

%%%%%%%%%%%%%%%%%%%%%%%%%%%%%%%%%%%%%%%%%%%%%%%%%%%%%%%%%%%%%%%%%%%%%%%%%%%%%%%%%%%%%%%%%%%%%%%%%%%%%%%%%%%%%%%%%%%%%%%%%%%%%%%%%%%%%%%%%%%%%%%%%%%%%%%%%%%%%%%%%%%%%%%%%%%%%%%%%%%%%%%%%%%%%%%%%%%%%%%%%%%%%%%%%%%%%%%%%%%%%%%%%%%%%%%%%%%%%%%%%%%%%%%%%%%%%%%%%%%%%%%%%%%%%%%%%%%%%%%%%%%%%%%%%%%%%%%%%%%%%%%%%%%%%%%%%%%%%%%%%%%%%%%%%%%%%%%%%%%%%%%%%%%%%%%%%%%%%%%%%%%%%%%%%%%%%%%%%%%%%%%%%%%%%%%%%%%%%%%%%%%%%%%%%%%%%%%%%%%%%%%%%%%%%%%%%%%%%%%%%%%%%%%%%%%%%%%%%%%%%%%%%%%%%%%%%%%%%%%%%%%%%%%%%%%%%%%%%%%%%%%%%%%%%%%%%%%%%%%%%%%%%%%%%%%%%%%%%%%%%%%%%%%%%%%%%%%%%%%%%%%%%%%%%%%%%%%%%%%%%%%%%%%%%%%%%%%%%%%%%%%%%%%%%%%%%%%%%%%%%%%%%%%%%%%%%%%%%%%%%%%%%%%%%%%%%%%%%%%%%%%%%%%%%%%%%%%%%%%%%%%%

\section{Conclusion}
\label{sec:conc}
In this work we have  computed from first principles the new supersymmetric index of a large class of torsional heterotic 
compactifications, corresponding to principal two-torus bundles over a warped K3 base. We have started with the worldsheet 
formulation of these flux compactifications as the IR fixed points of gauged linear sigma-models with torsion, and used 
supersymmetric localization methods. We have carefully explained why the localization procedure applies even in this context 
in which the classical action is non-invariant under the localization  supercharge. 

As an intermediate step of the computation we have defined a non-holomorphic genus $Z_{\textsc{fy}} (\tau, \bar \tau,y)$, 
that keeps track of the (non-holomorphic) $T^2$ zero-modes. For a generic Fu-Yau compactification, 
$i.e.$ with $\omega_1$ and $\omega_2$ non-collinear in the $H^2 (\mathcal{S},\mathbb{Z})$ lattice, this quantity is indeed invariant under continuous 
deformations of the theory, as the torus moduli $T$ and $U$ are then fully quantized by the flux. It is also, 
by construction, invariant under $F$-term and $D$-term deformations of the K3 base. 

We have given a geometrical formula for $Z_{\textsc{fy}} (\tau, \bar \tau,y)$, that provides a mathematical definition of 
this non-holomorphic genus independently of its realization as a path integral of a $(0,2)$ (non-)linear sigma-model. We have 
proven that this formula and the torsional GLSM result coincide in a simple example based 
on the quartic with a rank four holomorphic vector bundle. A proof of this equivalence for arbitrary models, 
which should follow from a direct generalization of the known results 
for complete intersection Calabi-Yau manifolds with the standard embedding, is an interesting mathematical project that is left for future work. 

Independently of physics, the non-holomorphic genus~(\ref{eq:nonholgen}) is of valuable mathematical interest. The elliptic genera 
of holomorphic gauge bundles of vanishing first Chern class over Calabi-Yau manifolds define Jacobi forms only 
if $\text{ch}_2 (\mathcal{T}_M) =\text{ch}_2 (\mathcal{V})$. In~\cite{Gritsenko:1999nm} a modified elliptic genus was defined 
by Gritsenko, in order to preserve this modular behavior even in particular when $\text{ch}_2 (\mathcal{T}_M) \neq \text{ch}_2 (\mathcal{V})$. In the present 
context there is an alternative definition motivated by physics; heterotic compactifications with 
$\text{ch}_2 (\mathcal{T}_M) \neq \text{ch}_2 (\mathcal{V})$ can be made anomaly-free if one adds an appropriate 
two-torus bundle over the Calabi-Yau manifold, leading naturally to the non-holomorphic genus~(\ref{eq:nonholgen}) transforming 
as a Jacobi form. 

This non-holomorphic genus  is presumably, as the CY elliptic genera, providing a generating functional for 
the indices of a family of Dirac operators, each transforming in a representation of the bundle specified by a given 
term in the expansion of~(\ref{eq:formalbundle}). A possible interpretation is that, in the present case, one considers a 
similar problem for Dirac operators related, in the string theory context, to Kaluza-Klein modes with momenta 
$(p_L,p_R)$ along the two-torus fiber, in their right Ramond ground state and with, roughly 
speaking, arbitrary left-moving oscillator modes along the tangent bundle of the base $\mathcal{T}_S$ and the 
gauge bundle $\mathcal{V}$. Because of the non-trivial fibration, one may expect a grading according to the toroidal left 
momentum $p_L$, as our explicit formula~(\ref{eq:nonholgen}) suggests. Making this correspondence more precise is a very 
interesting project.

As we have summarized in section~\ref{sec:indexdef}, the new supersymmetric index of $K3\times T^2$ compactifications is 
universal, hence the Mathieu moonshine conjecture, which postulates that the states of the $(4,4)$ theory underlying a $K3$ 
compactification with the standard embedding organize themselves into irreducible representations of the Mathieu group $M_{24}$, 
is apparent  regardless of the choice of gauge bundle~\cite{2013JHEP...09..030C}. 
In the present case, we don't expect that the result is universal, as for instance the choice of the 
complex two-form $\omega$ in the definition of  the Fu-Yau manifold affects the cohomology of the total space $\mathcal{M}$. 
An outstanding question is whether expanding the new supersymmetric index points towards $M_{24}$, or  a 
different sporadic group.

In physics, our results pave the way towards evaluating the one-loop  threshold 
corrections to the gauge and gravitational couplings in $\mathcal{N}=2$ heterotic compactifications with torsion. 
As the two-torus and K3 contributions are 
not factorized, the traditional 'unfolding method'~\cite{Dixon:1990pc} is likely not appropriate. Instead, one can use the 
new methods pioneered in~\cite{Angelantonj:2011br}, based on the Rankin-Selberg-Zagier approach to modular integrals. 
In this context a natural generalization of the present work is to extend our results to models which, together with the 
pullback of a holomorphic gauge bundle over the K3 surface, have  extra  Abelian bundles over the total space $\mathcal{M}$ 
(that would be   Wilson lines for $K3\times T^2$ compactifications). We plan to report on these topics in the next future.

Finally, it might be interesting to have a look at models with a non-compact target space. On the one hand, one could consider torus 
fibrations over a non-compact base, for instance over ALE and ALF spaces, and make the connection with the results obtained 
in~\cite{Carlevaro:2012rz} in the torsional case by one of the authors (were an exact worldsheet CFT description corresponding to  
an Eguchi-Hanson base was used), using a GLSM approach (see~\cite{Tong:2002rq,Harvey:2014nha} for examples without torsion). 
On the other hand,  one could define variants of the torsion GLSM in which one cannot reorganize the 
$(\mathbb{C}^{*})^{2}$ bundle into $\mathbb{C} \times T^2$, thereby giving torsional manifolds with non-compact fibers. 
In both cases we may observe some interesting features, such as an anomaly in the holomorphicity of the result, which would 
be given in terms of mock Jacobi forms (cf. for instance \cite{Ashok:2011cy,Murthy:2013mya,Harvey:2014nha}).

%%%%%%%%%%%%%%%%%%%%%%%%%%%%%%%%%%%%%%%%%%%%%%%%%%%%%%%%%%%%%%%%%%%%%%%%%%%%%%%%%%%%%%%%%%%%%%%%%%%%%%%%%%%%%%%%%%%%%%%%%%%%%%%%%%%%%%%%%%%%%%%%%%%%%%%%%%%%%%%%%%%%%%%%%%%%%%%%%%%%%%%%%%%%%%%%%%%%%%%%%%%%%%%%%%%%%%%%%%%%%%%%%%%%%%%%%%%%%%%%%%%%%%%%%%%%%%%%%%%%%%%%%%%%%%%%%%%%%%%%%%%%%%%%%%%%%%%%%%%%%%%%%%%%%%%%%%%%%%%%%%%%%%%%%%%%%%%%%%%%%%%%%%%%%%%%%%%%%%%%%%%%%%%%%%%%%%%%%%%%%%%%%%%%%%%%%%%%%%%%%%%%%%%%%%%%%%%%%%%%%%%%%%%%%%%%%%%%%%%%%%%%%%%%%%%%%%%%%%%%%%%%%%%%%%%%%%%%%%%%%%%%%%%%%%%%%%%%%%%%%%%%%%%%%%%%%%%%%%%%%%%%%%%%%%%%%%%%%%%%%%%%%%%%%%%%%%%%%%%%%%%%%%%%%%%%%%%%%%%%%%%%%%%%%%%%%%%%%%%%%%%%%%%%%%%%%%%%%%%%%%%%%%%%%%%%%%%%%%%%%%%%%%%%%%%%%%%%%%%%%%%%%%%%%%%%%%%%%%%%%%%%

\section*{Acknowledgments}

We thank Allan Adams, Luca Carlevaro, Miranda Cheng, Nima Doroud, Stefan Groot Nibbelink, Chris Hull and Jan Troost for 
discussions. We thank also the reviewer for useful comments and suggestions. D.I. thanks the DAMTP of Cambridge University for its hospitality while part of this research  
was done. This work was conducted within the ILP LABEX (ANR-10-LABX-63) supported by French state funds managed by the 
ANR (ANR-11-IDEX-0004-02) and by the project QHNS in the program ANR Blanc 
SIMI5 of Agence National de la Recherche. 

%%%%%%%%%%%%%%%%%%%%%%%%%%%%%%%%%%%%%%%%%%%%%%%%%%%%%%%%%%%%%%%%%%%%%%%%%%%%%%%%%%%%%%%%%%%%%%%%%%%%%%%%%%%%%%%%%%%%%%%%%%%%%%%%%%%%%%%%%%%%%%%%%%%%%%%%%%%%%%%%%%%%%%%%%%%%%%%%%%%%%%%%%%%%%%%%%%%%%%%%%%%%%%%%%%%%%%%%%%%%%%%%%%%%%%%%%%%%%%%%%%%%%%%%%%%%%%%%%%%%%%%%%%%%%%%%%%%%%%%%%%%%%%%%%%%%%%%%%%%%%%%%%%%%%%%%%%%%%%%%%%%%%%%%%%%%%%%%%%%%%%%%%%%%%%%%%%%%%%%%%%%%%%%%%%%%%%%%%%%%%%%%%%%%%%%%%%%%%%%%%%%%%%%%%%%%%%%%%%%%%%%%%%%%%%%%%%%%%%%%%%%%%%%%%%%%%%%%%%%%%%%%%%%%%%%%%%%%%%%%%%%%%%%%%%%%%%%%%%%%%%%%%%%%%%%%%%%%%%%%%%%%%%%%%%%%%%%%%%%%%%%%%%%%%%%%%%%%%%%%%%%%%%%%%%%%%%%%%%%%%%%%%%%%%%%%%%%%%%%%%%%%%%%%%%%%%%%%%%%%%%%%%%%%%%%%%%%%%%%%%%%%%%%%%%%%%%%%%%%%%%%%%%%%%%%%%%%%%%%%%%%%

\appendix

\section{\texorpdfstring{$\mathbf{(0,2)}$}{(0,2)} superspace and Lagrangians\label{appendixConventions}}

	In this appendix we  summarize the field content and interactions of two-dimensional field theories with $(0,2)$ supersymmetry. We  denote by $(\sigma^{0},\sigma^{1})$ the Lorentzian coordinates; the analytic continuation to Euclidean signature is obtained by 
	defining $\sigma^{2}=-i\sigma^{0}$, and considering the complex variables\footnote{In our conventions  left-moving$\ \leftrightarrow\ $holomorphic.} 
	$z=\sigma^{1}+i\sigma^{2}$, $\bar{z}=\sigma^{1}-i\sigma^{2}$. Then Euclidean $(0,2)$ superspace is spanned by the coordinates 
	$(z,\bar z,\theta^+,\theta^-)$. We define the superspace covariant derivatives as:
	\begin{equation}
	\bar{D}_\pm = \frac{\partial}{\partial \theta^{\pm}}+ \theta^{\mp} \bar \partial\, .
	\end{equation}

	The matter part of the  theory is made of two types of multiplets. First, chiral multiplets correspond to 
	$(0,2)$ superfields satisfying the condition $\bar{D}_{+}\Phi=0$, hence of component expansion 
	\begin{equation}
		\Phi=\phi+\theta^{+}\psi_{+}+\theta^{+}\theta^{-}\bar{\partial}\phi\, ,
	\end{equation} 
	composed of a complex scalar and a right-moving Weyl fermion. Second, one defines Fermi mutiplets also satisfying\footnote{There exists a generalization where the right hand-side, instead of being zero, consists of a holomorphic function $E(\Phi)$ of the chiral multiplets of the theory.} $\bar{D}_{+}\Lambda=0$, whose bottom component is a left-moving Weyl fermion, with components expansion 
	\begin{equation}
		\Lambda=\gamma_{-}+\theta^{+}G+\theta^{+}\theta^{-}\bar{\partial}\gamma_{-}\, , 
	\end{equation}
	where $G$ is an auxiliary field. 
		
Gauge interactions are mediated by a gauge multiplet which corresponds, in $(0,2)$ superspace, to a pair of superfields $(A_+,A_-)$. 
In Wess-Zumino gauge, the component expansion reads:
\begin{subequations}
	\begin{align}
		A_{-}&=a_{z}+\theta^{-}\lambda_{-}-\theta^{+}\bar{\lambda}_{-}+\frac{1}{2}\theta^{+}\theta^{-}D\, ,\\
		A_{+}&=\theta^{+}\theta^{-}a_{\bar{z}}\, ,
	\end{align}
with $(\lambda_{-},\bar{\lambda}_-)$ the gaugini, and $D$ the auxiliary superfield. Under a supergauge transformation of chiral superfield parameter $\Xi$, one has $\delta_{\Xi}A_{-}=-i\left(\Xi+\bar{\Xi}\right)$ and $\delta_{\Xi}A_{+}=-i\left(\Xi-\bar{\Xi}\right)$. The gauge superfield strength is a chiral superfield, given by 
	\begin{equation}
		\Upsilon=-\lambda_{-}+\theta^{+}(D+a_{z\bar{z}})-\theta^{+}\theta^{-}\bar{\partial}\lambda_{-}\, .
	\end{equation}
\end{subequations}

	Interactions are then specified by minimally coupling the matter superfields to the gauge superfields, and by a set of holomorphic functions $J^{\alpha}(\Phi)$ of the chiral multiplets, one for each Fermi multiplet in the theory; if the gauge 
	group contains also $U(1)$ factors, one can add Fayet-Iliopoulos terms. The full Lagrangian density then writes in terms of components
	\begin{equation}											   
		\mathcal{L}=\mathcal{L}_{\text{c.m.}}+\mathcal{L}_{\text{f.m}}+\mathcal{L}_{\text{v.m.}}+\mathcal{L}_{\textsc{j}}+\mathcal{L}_{\textsc{fi}}\, ,
	\end{equation}
	with
\begin{subequations}
	\begin{align}
		\mathcal{L}_{\text{c.m.}}&=\nabla_{z}\bar{\phi}\nabla_{\bar{z}}\phi+\nabla_{\bar{z}}\bar{\phi}\nabla_{z}\phi+2\bar{\psi}\nabla_{z}\psi-Q\lambda\psi\bar{\phi}+Q\bar{\lambda}\bar{\psi}\phi-QD|\phi|^{2}\, ,\\
		\mathcal{L}_{\text{f.m.}}&=2\bar{\gamma}\nabla_{\bar{z}}\gamma+|G|^{2}\, ,\\
		\mathcal{L}_{\text{v.m.}}&=\frac{1}{2e^{2}}\left(a_{z\bar{z}}^{\ 2}+2\bar{\lambda}\bar{\partial}\lambda-D^{2}\right)\, ,\\
		\mathcal{L}_{\textsc{j}}&=G_{\alpha}J^{\alpha}-\gamma_{\alpha}\psi^{i}\partial_{i}J^{\alpha}+\text{h.c.}\, ,\\
		\mathcal{L}_{\textsc{fi}}&=\xi \, D-\frac{i\theta}{2\pi}a_{z\bar{z}}\, ,
	\end{align}
\end{subequations}
	where $a_{z\bar{z}}=2\left(\partial a_{\bar{z}}-\bar{\partial}a_{z}\right)$ is the gauge field strength.
		
	In the Wess-Zumino gauge, supersymmetry transformations should be followed by a supergauge transformation of chiral superfield parameter $\Xi_{wz}=i\bar{\epsilon}\theta^{+}a_{\bar{z}}$ in order 
	to restore the gauge choice. Under the full transformation, defined as $\delta_{\epsilon}=\left(\epsilon Q_{+}-\bar{\epsilon}\bar{Q}_{+}+\delta_{\text{gauge}}\right)$, 
	the component fields of the various multiplets, including the torsion multiplet, behave as	
	\hfill
	\begin{minipage}[c]{.22\linewidth}
		\begin{align}
			&\delta_{\epsilon}\phi=\bar{\epsilon}\psi\nonumber\\
			&\delta_{\epsilon}\bar{\phi}=-\epsilon\bar{\psi}\nonumber\\
			&\delta_{\epsilon}\psi=\epsilon\nabla_{\bar{z}}\phi\nonumber\\
			&\delta_{\epsilon}\bar{\psi}=-\bar{\epsilon}\nabla_{\bar{z}}\bar{\phi}\nonumber\\
			&\delta_{\epsilon}\gamma=-\bar{\epsilon}G\nonumber\\
			&\delta_{\epsilon}\bar{\gamma}=-\epsilon\bar{G}\nonumber
		\end{align}
	\end{minipage} \hfill
	\begin{minipage}[c]{.25\linewidth}
		\begin{align}
			&\delta_{\epsilon} G=2\epsilon\nabla_{\bar{z}}\gamma\nonumber\\
			&\delta_{\epsilon}\bar{G}=2\bar{\epsilon}\nabla_{\bar{z}}\bar{\gamma}\nonumber\\
			&\delta_{\epsilon}\alpha=\bar{\epsilon}\chi\nonumber\\
			&\delta_{\epsilon}\bar{\alpha}=-\epsilon\bar{\chi}\nonumber\\
			&\delta_{\epsilon}\chi=\epsilon\nabla_{\bar{z}}\alpha\nonumber\\
			&\delta_{\epsilon}\bar{\chi}=-\bar{\epsilon}\nabla_{\bar{z}}\bar{\alpha}\nonumber
		\end{align}
	\end{minipage} \hfill
	\begin{minipage}[c]{.30\linewidth}
		\begin{align}
			&\delta_{\epsilon} a_{z}=\frac{1}{2}\left(\epsilon\bar{\lambda}+\bar{\epsilon}\lambda\right)\nonumber\\
			&\delta_{\epsilon} a_{\bar{z}}=0\nonumber\\
			&\delta_{\epsilon}\lambda=\epsilon\left(a_{z\bar{z}}+D\right)\ \label{eq:comptrans}  \\
			&\delta_{\epsilon}\bar{\lambda}=\bar{\epsilon}\left(a_{z\bar{z}}-D\right)\nonumber\\
			&\delta_{\epsilon} D=-\left(\epsilon\bar{\partial}\bar{\lambda}-\bar{\epsilon}\bar{\partial}\lambda\right)\nonumber
\end{align}
	\end{minipage} 
	\hfill
		
%%%%%%%%%%%%%%%%%%%%%%%%%%%%%%%%%%%%%%%%%%%%%%%%%%%%%%%%%%%%%%%%%%%%%%%%%%%%%%%%%%%%%%%%%%%%%%%%%%%%%%%%%%%%%%%%%%%%%%%%%%%%%%%%%%%%%%%%%%%%%%%%%%%%%%%%%%%%%%%%%%%%%%%%%%%%%%%%%%%%%%%%%%%%%%%%%%%%%%%%%%%%%%%%%%%%%%%%%%%%%%%%%%%%%%%%%%%%%%%%%%%%%%%%%%%%%%%%%%%%%%%%%%%%%%%%%%%%%%%%%%%%%%%%%%%%%%%%%%%%%%%%%%%%%%%%%%%%%%%%%%%%%%%%%%%%%%%%%%%%%%%%%%%%%%%%%%%%%%%%%%%%%%%%%%%%%%%%%%%%%%%%%%%%%%%%%%%%%%%%%%%%%%%%%%%%%%%%%%%%%%%%%%%%%%%%%%%%%%%%%%%%%%%%%%%%%%%%%%%%%%%%%%%%%%%%%%%%%%%%%%%%%%%%%%%%%%%%%%%%%%%%%%%%%%%%%%%%%%%%%%%%%%%%%%%%%%%%%%%%%%%%%%%%%%%%%%%%%%%%%%%%%%%%%%%%%%%%%%%%%%%%%%%%%%%%%%%%%%%%%%%%%%%%%%%%%%%%%%%%%%%%%%%%%%%%%%%%%%%%%%%%%%%%%%%%%%%%%%%%%%%%%%%%%%%%%%%%%%%%%%%%

\section{Poisson resummation formula, theta functions and Eisenstein series\label{appendixTheta}}
\label{app:theta}
	
	For a general n-dimensional lattice $\Gamma$, with $A$ a symmetric positive definite $n\times n$ matrix defining its bilinear form, one has the Poisson resummation formula
	\begin{equation}
		\sum_{p\in\Gamma}e^{-\pi(p+x)\cdot A(p+x)+2i\pi y\cdot(p+x)}=\frac{1}{\text{vol}(\Gamma)\sqrt{\text{det}A}}\sum_{p\in\Gamma^{\vee}}e^{-2i\pi p\cdot x-\pi(y+p)\cdot A^{-1}(y+p)}\, ,
	\end{equation}
	which reduces in the case $\Gamma=\mathbb{Z}$ to
	\begin{equation}
		\sum_{n=-\infty}^{\infty}e^{-\pi an^{2}+2i\pi bn}=\frac{1}{\sqrt{a}}\sum_{n=-\infty}^{\infty}e^{-\frac{\pi}{a}(n-b)^{2}}\, .
	\end{equation}
	The Dedekind eta function is defined by
	\begin{equation}
		\eta(\tau)=q^{\frac{1}{24}}\prod_{n=1}^{\infty}\left(1-q^{n}\right)\, ,
	\end{equation}
	with $q=e^{2i\pi\tau}$. Its modular properties are
\begin{subequations}
	\begin{align}
		\eta(\tau+1)&=e^{i\frac{\pi}{12}}\eta(\tau)\, ,\\
		\eta\left(-\frac{1}{\tau}\right)&=(-i\tau)^{1/2}\eta(\tau)\, .
	\end{align}
\end{subequations}
	The Jacobi theta functions with characteristics are defined by
	\begin{equation}
		\vartheta\left[^{a}_{b}\right](\tau|u)=\sum_{n\in\mathbb{Z}}q^{\frac{1}{2}(n+\frac{a}{2})^{2}}e^{2i\pi(n+\frac{a}{2})(u+\frac{b}{2})}\, .
	\end{equation}
	One defines 
	\begin{equation}
		\vartheta_{1}=-\vartheta\left[^{1}_{1}\right]\ \ \ \vartheta_{2}=\vartheta\left[^{1}_{0}\right]\ \ \ \vartheta_{3}=\vartheta\left[^{0}_{0}\right]\ \ \ \vartheta_{4}=\vartheta\left[^{0}_{1}\right]\, .
	\end{equation}
	One can rewrite the Jacobi theta functions in terms of an infinite product. In particular, $\vartheta_{1}$ writes:
	\begin{equation}
		\vartheta_{1}(\tau|u)=-iq^{\frac{1}{8}}w^{\frac{1}{2}}\prod_{n=1}^{\infty}\left(1-q^{n}\right)\left(1-wq^{n}\right)\left(1-w^{-1}q^{n-1}\right)\, ,
	\end{equation}
	with $w:=\exp(2i\pi u)$.
	One has the following properties
\begin{subequations}
	\begin{align}
		&\frac{\partial}{\partial u}\vartheta_{1}(\tau|u)|_{u=0}=2\pi\eta(\tau)^{3}\, ,\label{propTheta}\\
		&\vartheta_{1}(\tau|-u)=-\vartheta_{1}(\tau|u)\, ,\\
		&\oint_{u=k+\tau l}\frac{\d u}{2i\pi}\ \frac{1}{\vartheta_{1}(\tau|u)}=i\frac{(-1)^{k+l}e^{i\pi l^{2}\tau}}{\eta(\tau)^{3}}\, .
	\end{align}
\end{subequations}
	Under modular transformations, the Jacobi theta functions transform as
	\begin{subequations}
		\begin{align}
			\vartheta\left[^{a}_{b}\right]\left(\tau+1|u\right)&=e^{-\frac{i\pi}{4}a(a-2)}\theta\left[^{\ \ \ a}_{a+b-1}\right]\left(\tau|u\right)\, ,\\
			\vartheta\left[^{a}_{b}\right]\left(\left.-\frac{1}{\tau}\right|\frac{u}{\tau}\right)&=-\sqrt{-i\tau}e^{\frac{i\pi}{2}ab+\frac{i\pi u^{2}}{\tau}}\theta\left[^{\ b}_{-a}\right]\left(\tau|u\right)\, .
		\end{align}
	\end{subequations}
	They satisfy the quasi-periodicity property
	\begin{equation}
		\vartheta\left[^{a}_{b}\right](\tau|u+m+\tau n)=\exp\left(i\pi ma-i\pi\tau n^{2}-2i\pi n\left(u+\frac{b}{2}\right)\right)\vartheta\left[^{a}_{b}\right](\tau|u)\ \ \ \ \text{for $m,n\in\mathbb{Z}$}\, .\label{eq:periodicity1}
	\end{equation}
	The $\widehat{\mathfrak{su}(2)}_{k}$ theta functions are defined by
	\begin{equation}
		\Theta_{s,k}(\tau|z)=\sum_{n\in\mathbb{Z}+\frac{s}{2k}}q^{kn^{2}}e^{2i\pi zkn}\, ,
	\end{equation}
	with $s\in\mathbb{Z}_{2k}$.
	Under modular transformations, one has
	\begin{subequations}
		\begin{align}
			\Theta_{s,k}(\tau+1|z)&=e^{i\pi\frac{s^{2}}{k}}\Theta_{s,k}(\tau|z)\, ,\\
			\Theta_{s,k}\left(\left.-\frac{1}{\tau}\right|\frac{z}{\tau}\right)&=(-i\tau)^{1/2}\frac{1}{\sqrt{2k}}e^{\frac{i\pi}{\tau}\frac{kz^{2}}{2}}\sum_{s'\in\mathbb{Z}_{2k}}e^{-\frac{i\pi}{k}ss'}\Theta_{s',k}(\tau|z)\, .
		\end{align}
	\end{subequations}
	They also satisfy a quasi-periodicity property
	\begin{equation}
		\Theta_{s,k}(\tau|z+m+\tau n)=(-1)^{k(m+n)}e^{-i\pi k\left(\frac{n^{2}}{2}\tau+nz\right)}\Theta_{s,k}(\tau|z)\, .\label{eq:periodicity2}
	\end{equation}
	Following \cite{Kac:1984mq}, one defines the theta function related to a lattice $\Gamma$ by 
	\begin{equation}
		\Theta_{\mu}^{\Gamma}(\tau|\lambda)=\sum_{\gamma\in\Gamma+\mu}q^{\frac{1}{2}\langle\gamma,\gamma\rangle}e^{2i\pi\langle\gamma,\lambda\rangle}\, .
	\end{equation}
	Under modular transformations, one has
	\begin{subequations}
		\begin{align}
			\Theta_{\mu}^{\Gamma}(\tau+1|\lambda)&=e^{i\pi\langle\mu,\mu\rangle}\Theta_{\mu}^{\Gamma}(\tau|\lambda)\, ,
			\label{eq:thetatransA}\\
			\Theta_{\mu}^{\Gamma}\left(-\left.\frac{1}{\tau}\right|\frac{\lambda}{\tau}\right)&=\frac{\left(-i\tau\right)^{\frac{\text{rank}(\Gamma)}{2}}}{|\Gamma^{\vee}/\Gamma|^{\frac{1}{2}}}e^{\frac{i\pi}{\tau}\langle\lambda\, ,\lambda\rangle}\sum_{\mu'\in\Gamma^{\vee}/\Gamma}e^{-2i\pi\langle\mu,\mu'\rangle}\Theta_{\mu'}^{\Gamma}(\tau|\lambda)\, .
		\end{align}
	\label{eq:thetatrans}
	\end{subequations}
	Let us define the Kronecker delta on the lattice $\Gamma$ by:
	\begin{equation}
	\label{eq:knonecker}
		\delta_{b,b'}=\frac{1}{|\Gamma^{\vee}/\Gamma|}\sum_{a\in\Gamma^{\vee}/\Gamma}e^{2i\pi\langle a,b-b'\rangle}\, .
	\end{equation}
	Given a triplet $(\Gamma_{\textsc{l}},\Gamma_{\textsc{r}},\varphi)$, with $\varphi$ being an isometry between the discriminant group of the two lattices $\Gamma_{\textsc{l}}$ and $\Gamma_{\textsc{r}}$, let us determine the modular behaviour under a S-transformation of the quantity
	\begin{equation}
		\sum_{\mu\in\Gamma_{\textsc{l}}^{\vee}/\Gamma_{\textsc{l}}}\Theta_{\mu}^{\Gamma_{\textsc{l}}}\left(\left.\tau\right|\lambda\right)\bar{\Theta}_{\varphi(\mu)}^{\Gamma_{\textsc{r}}}\left(\left.\bar{\tau}\right|0\right)\, .
	\end{equation}
	One has
	\begin{align}
		\sum_{\mu\in\Gamma_{\textsc{l}}^{\vee}/\Gamma_{\textsc{l}}}\Theta_{\mu}^{\Gamma_{\textsc{l}}}\left(\left.-\frac{1}{\tau}\right|\frac{\lambda}{\tau}\right)&\bar{\Theta}_{\varphi(\mu)}^{\Gamma_{\textsc{r}}}\left(\left.-\frac{1}{\bar{\tau}}\right|0\right)=\nonumber\\
		&\frac{|\tau|^{\text{rank}(\Gamma)}}{|\Gamma_{\textsc{l}}^{\vee}/\Gamma_{\textsc{l}}|}e^{\frac{i\pi}{\tau}\langle\lambda,\lambda\rangle}\sum_{\mu}\sum_{\rho,\bar{\rho}}e^{-2i\pi\left(\langle\mu,\rho\rangle-\langle\varphi(\mu),\bar{\rho}\rangle\right)}\Theta_{\rho}^{\Gamma_{\textsc{l}}}\left(\left.\tau\right|\lambda\right)\bar{\Theta}_{\bar{\rho}}^{\Gamma_{\textsc{r}}}\left(\left.\bar{\tau}\right|0\right)\nonumber\\
	\end{align}
	Since $\varphi$ is an isometry, one has $\langle\varphi(\mu),\bar{\rho}\rangle=\langle\mu,\varphi^{-1}(\bar{\rho})\rangle$. Permuting the sums, using eq. (\ref{eq:knonecker}), and the fact that $\delta_{\rho,\varphi^{-1}(\bar{\rho})}=\delta_{\bar{\rho},\varphi(\rho)}$, one obtains finally
	\begin{equation}
		\sum_{\mu\in\Gamma_{\textsc{l}}^{\vee}/\Gamma_{\textsc{l}}}\Theta_{\mu}^{\Gamma_{\textsc{l}}}\left(\left.-\frac{1}{\tau}\right|\frac{\lambda}{\tau}\right)\bar{\Theta}_{\varphi(\mu)}^{\Gamma_{\textsc{r}}}\left(\left.-\frac{1}{\bar{\tau}}\right|0\right)=|\tau|^{\text{rank}(\Gamma)}e^{\frac{i\pi}{\tau}\langle\lambda,\lambda\rangle}\sum_{\mu\in\Gamma_{\textsc{l}}^{\vee}/\Gamma_{\textsc{l}}}\Theta_{\mu}^{\Gamma_{\textsc{l}}}\left(\left.\tau\right|\lambda\right)\bar{\Theta}_{\varphi(\mu)}^{\Gamma_{\textsc{r}}}\left(\left.\bar{\tau}\right|0\right)\, .
	\end{equation}
	The weight $2k$ ($k>1$) Eisenstein series are holomorphic modular forms given by
	\begin{equation}
		E_{2k}(\tau)=-\frac{(2k)!}{(2i\pi)^{2k}B_{2k}}\sum_{(m,n)\neq (0,0)}\frac{1}{(m\tau+n)^{2k}}\, ,
	\end{equation}
	with $B_{2k}$ the Bernoulli numbers. In terms of Jacobi theta functions, one has
	\begin{equation}
		E_{4}(\tau)=\frac{1}{2}\left(\vartheta_{2}(\tau)^{8}+\vartheta_{3}(\tau)^{8}+\vartheta_{4}(\tau)^{8}\right)\, .
	\end{equation}
	In particular, $E_{4}/\eta^{8}$ is modular invariant.
	Finally, a weak Jacobi form of weight $k$ and index $t$ with 
	character $\chi$ is a holomorphic function $\phi$ on $\mathbb{H}\times \mathbb{C}$ which satisfies 
	\begin{subequations}
		\begin{align}
			\phi \left( \frac{a\tau +b}{c\tau +d} ,\frac{y}{c\tau +d} \right)  
			= \chi \,  \left( g \right)\, 
			(c\tau +d)^k e^{2i\pi t c \frac{y^2}{c\tau + d}}\ \phi (\tau,y) \quad &, \qquad 
			g= \left( \begin{array}{cc} a &b\\c &d \end{array}\right) \in SL(2,\mathbb{Z})\, ,\\
			\phi (\tau,y + \lambda \tau + \mu ) = (-1)^{2t(\lambda + \mu)}e^{-2i\pi t(\lambda^2 t + 2 \lambda y)} \ 
			\phi (\tau, y) \quad &, \qquad \lambda, \mu \in \mathbb{Z}\, ,
		\end{align}
	\end{subequations}
	and with a Fourier expansion containing only positive powers of $q$. 

%%%%%%%%%%%%%%%%%%%%%%%%%%%%%%%%%%%%%%%%%%%%%%%%%%%%%%%%%%%%%%%%%%%%%%%%%%%%%%%%%%%%%%%%%%%%%%%%%%%%%%%%%%%%%%%%%%%%%%%%%%%%%%%%%%%%%%%%%%%%%%%%%%%%%%%%%%%%%%%%%%%%%%%%%%%%%%%%%%%%%%%%%%%%%%%%%%%%%%%%%%%%%%%%%%%%%%%%%%%%%%%%%%%%%%%%%%%%%%%%%%%%%%%%%%%%%%%%%%%%%%%%%%%%%%%%%%%%%%%%%%%%%%%%%%%%%%%%%%%%%%%%%%%%%%%%%%%%%%%%%%%%%%%%%%%%%%%%%%%%%%%%%%%%%%%%%%%%%%%%%%%%%%%%%%%%%%%%%%%%%%%%%%%%%%%%%%%%%%%%%%%%%%%%%%%%%%%%%%%%%%%%%%%%%%%%%%%%%%%%%%%%%%%%%%%%%%%%%%%%%%%%%%%%%%%%%%%%%%%%%%%%%%%%%%%%%%%%%%%%%%%%%%%%%%%%%%%%%%%%%%%%%%%%%%%%%%%%%%%%%%%%%%%%%%%%%%%%%%%%%%%%%%%%%%%%%%%%%%%%%%%%%%%%%%%%%%%%%%%%%%%%%%%%%%%%%%%%%%%%%%%%%%%%%%%%%%%%%%%%%%%%%%%%%%%%%%%%%%%%%%%%%%%%%%%%%%%%%%%%%%%%

\section{Fu-Yau compactifications\label{appendixFuYau}}

	Strominger established in \cite{Strominger:1986uh} the conditions that a heterotic compactification to four dimensions with $\mathcal{N}=1$ supersymmetry should satisfy. The internal manifold should be a complex manifold with $SU(3)$ structure, which is 
	characterized by a holomorphic $(3,0)$-form $\Omega$ and a  Hermitian $(1,1)$-form $J$,  together with the field strengh of a 
	connection on a holomorphic gauge bundle $V$, subject to the following system of equations, known as Strominger's system:
	\begin{subequations}
		\begin{align}
			&\d \left(||\Omega||_{J}J\wedge J\right)=0\, ,\label{1}\\
			&F_{ij}=F_{\bar{\imath}\bar{\jmath}}=0\, ,\label{2}\\
			&J^{i\bar \jmath}F_{i\bar \jmath}=0\, ,\label{3}\\
			&\d H= 2i\partial\bar{\partial}J=\frac{\alpha'}{4}\left(\text{tr}(R\wedge R)-\text{tr}(F\wedge F)\right)\label{4}\, ,
		\end{align}
	\end{subequations}
	where $||\star||_{J}$ is the norm corresponding to the hermitian scalar product $(\star,\star)_{J}$ defined with respect to the fundamental form $J$.
	The physical fields, i.e. metric, torsion and dilaton field are then expressed in terms the $\Omega$ and $J$, solutions to this system. 

	Equation \eqref{1} means that the internal manifold should be conformally balanced, which is weaker than K\"ahlerity. Equations \eqref{2},\eqref{3} constraining the gauge bundle are the 
	zero-slope limit of the Hermitian-Yang-Mills equations, which can be rephrased as demanding that the vector bundle be a holomorphic stable vector bundle. 
	Finally, equation \eqref{4} is the generalized Bianchi identity, a consequence of the Green-Schwarz mechanism. The right-hand side of~\eqref{4} is computed using a connection 
	with torsion on the tangent bundle; various possible choices correspond to different regularization schemes in the underlying non-linear sigma-model~\cite{Hull:1985dx}. 
		
	Solutions of this system with non-zero $H$-flux, corresponding to a principal $T^{2}$-bundle $\mathcal{M}$ over a warped K3 base (which we denote $\mathcal{S}$), $T^{2}\hookrightarrow\mathcal{M}\overset{\pi}{\rightarrow}\mathcal{S}$, were first obtained in~\cite{Dasgupta:1999ss} through string dualities. Then the underlying $SU(3)$ structure was studied by Goldstein and Prokushkin in~\cite{Goldstein:2002pg}, and finally Fu and Yau obtained a solution of the generalized Bianchi identity~\eqref{4} using the Hermitian connection on the tangent bundle in~\cite{Fu:2006vj}. Following the common usage, these solutions will be called Fu-Yau compactifications. 

	Explicitly, taking a two-torus of moduli $T$ and $U$, see eq.~(\ref{eq:torusmod}), the metric on the internal six-dimensional manifold $\mathcal{M}$ is chosen to be of the form 
	\begin{equation}
		\label{metric}
		\d s^2 = 
		\frac{U_2}{T_2} \left|\d x^1 + T \d x^2 +  \pi^\star \alpha \right|^2 
		+ e^{2 A(y)} \d s^2 (\mathcal{S})\, ,
	\end{equation}
	where $\d s^2 (\mathcal{S})$ is a Ricci-flat metric on a K3 surface $\mathcal{S}$ and $e^{2A}$ is 
	a warp factor depending on the K3 coordinates only. The connection one-form $\alpha$ on  $\mathcal{S}$ 
	is such that 
	\begin{equation}
		\iota=\d x^1 + T \d x^2 +  \pi^\star \alpha
	\end{equation}
	is a globally defined $(1,0)$ form on $\mathcal{M}$. We then define the complex two-form $\omega$ on $\mathcal{S}$  through
	\begin{equation}
		\tfrac{1}{2\pi}\, \d \iota =   \pi^\star \omega\, ,
	\end{equation}
	that we expand in terms of the $T^2$ complex structure as 
	\begin{equation}
		\omega = \omega_1 + T \omega_2 \, .
	\end{equation}
	The metric~(\ref{metric}) is globally defined provided  that $\omega_\ell\in H^2(\mathcal{S},\mathbb{Z})$. 

	As was shown by Goldstein and Prokushkin, a solution of the supersymmetry conditions is obtained provided that 
	$\omega$ has no component in $\Lambda^{0,2} T^\star \mathcal{S}$ and is primitive, $i.e.$ such that
	\begin{equation}
		\omega\wedge J_{\mathcal{S}}=0\, .
		\label{eq:primitivity}
	\end{equation}
	One can then obtain the $(1,1)$ form $J$ and the $(3,0)$ form $\Omega$ 
	characterizing the $SU(3)$ structure  in terms of the K\"ahler form and 
	holomorphic two-form on $\mathcal{S}$, $J_\mathcal{S}$ and $\Omega_\mathcal{S}$, as 
	\begin{subequations}
		\begin{align}
			&\Omega=\pi^{*}\left(\Omega_{\mathcal{S}}\right)\wedge\iota\, ,\\
			&J=\pi^{*}\left(e^{2\phi}J_{\mathcal{S}}\right)+\frac{iU_2}{2T_{2}}\iota\wedge\bar{\iota}\, .
		\end{align}
	\end{subequations}

	Solutions with extended $\mathcal{N}=2$ supersymmetry in four dimensions, 
	$i.e.$ with $SU(2)$ structure, are obtained imposing the extra condition $\omega \in H^{(1,1)} (\mathcal{S})$. This is the relevant 
	case for the torsion gauged linear sigma-models that we consider in this work. 

	One can for instance choose $\omega_{1,2}$  in the Picard lattice 
	of $\mathcal{S}$, defined by  $\text{Pic}(\mathcal{S})=H^{2} (\mathcal{S},\mathbb{Z}) \cap  H^{(1,1)} (\mathcal{S})$, whose 
	rank is denoted  $\rho(\mathcal{S})$. Let us  define a set of complex topological charges 
	$\{ M^n, n = 1,\ldots,\rho(\mathcal{S})\}$, belonging to the lattice $\mathbb{Z} + T \, \mathbb{Z}$, and choose 
	a basis of $\text{Pic}(\mathcal{S})$, $\{ \varpi_n , n=1,\ldots,\rho(\mathcal{S}) \}$. One expands the curvature of the two-torus bundle as 
	\begin{equation}
	\label{eq:omegaexpansion}
		\omega =\sum_{n=1}^{\rho(\mathcal{S})} M^n \varpi_n\, .
	\end{equation}

	The vector bundle over $\mathcal{M}$ is obtained as the pullback of a holomorphic gauge bundle $V$ on K3 satisfying the zero-slope 
	limit of the Hermitian-Yang-Mills equations, see eqns.~(\ref{2},\ref{3}). On K3 it implies anti-self-duality, $i.e.$ that the bundle $V$ corresponds to an anti-instanton background.   Fu and Yau showed in~\cite{Fu:2006vj} that one can find a smooth solution to the Bianchi identity for the warp factor, using the Chern connection, provided the following tadpole condition holds,  
	\begin{equation}
		\int_{\mathcal{S}}\text{ch}_{2}(V)+\frac{U_{2}}{T_{2}}d_{mn} M^m \bar{M}^n+24=0\, ,
	\label{eq:tadpole_condition}
	\end{equation}
	written the basis~(\ref{eq:omegaexpansion}), where  $d_{mn}$ is the intersection matrix on $H^{2} (\mathcal{S},\mathbb{Z})$. 

%%%%%%%%%%%%%%%%%%%%%%%%%%%%%%%%%%%%%%%%%%%%%%%%%%%%%%%%%%%%%%%%%%%%%%%%%%%%%%%%%%%%%%%%%%%%%%%%%%%%%%%%%%%%%%%%%%%%%%%%%%%%%%%%%%%%%%%%%%%%%%%%%%%%%%%%%%%%%%%%%%%%%%%%%%%%%%%%%%%%%%%%%%%%%%%%%%%%%%%%%%%%%%%%%%%%%%%%%%%%%%%%%%%%%%%%%%%%%%%%%%%%%%%%%%%%%%%%%%%%%%%%%%%%%%%%%%%%%%%%%%%%%%%%%%%%%%%%%%%%%%%%%%%%%%%%%%%%%%%%%%%%%%%%%%%%%%%%%%%%%%%%%%%%%%%%%%%%%%%%%%%%%%%%%%%%%%%%%%%%%%%%%%%%%%%%%%%%%%%%%%%%%%%%%%%%%%%%%%%%%%%%%%%%%%%%%%%%%%%%%%%%%%%%%%%%%%%%%%%%%%%%%%%%%%%%%%%%%%%%%%%%%%%%%%%%%%%%%%%%%%%%%%%%%%%%%%%%%%%%%%%%%%%%%%%%%%%%%%%%%%%%%%%%%%%%%%%%%%%%%%%%%%%%%%%%%%%%%%%%%%%%%%%%%%%%%%%%%%%%%%%%%%%%%%%%%%%%%%%%%%%%%%%%%%%%%%%%%%%%%%%%%%%%%%%%%%%%%%%%%%%%%%%%%%%%%%%%%%%%%%%%

\bibliographystyle{JHEP}
\bibliography{biblioJ}
%\nocite{*}

\end{document}